\documentclass{article}

\usepackage{arxiv}
\usepackage[sort]{cite}

\usepackage[utf8]{inputenc} 
\usepackage[T1]{fontenc}    
\usepackage{url}            
\usepackage{booktabs}       
\usepackage{amsfonts}       
\usepackage{nicefrac}       
\usepackage{microtype}      
\usepackage{lipsum}
\usepackage{balance}
\usepackage{algorithmic}
\usepackage[ruled, lined]{algorithm2e} 
\usepackage{graphicx}
\usepackage{textcomp}
\usepackage{xcolor}
\usepackage{enumitem} 
\usepackage{float}
\usepackage{tabularx}
\usepackage{threeparttable} 
\usepackage{stfloats}   
\usepackage{placeins}   
\usepackage{balance}

\usepackage{cuted} 
\usepackage{tcolorbox}
\usepackage{amsthm} 
\usepackage{amsmath}  

\usepackage{array}
\usepackage{multirow}
\usepackage{colortbl}
\usepackage{placeins}
\usepackage{float}
\usepackage{tabulary}
\usepackage{lscape}
\usepackage{rotating}
\usepackage{makecell}
\usepackage[usestackEOL]{stackengine}
\usepackage{longtable}
\usepackage[font=small,format=plain,labelfont=bf,textfont=it]{caption} 
\usepackage{graphicx}
\graphicspath{ {./figures/} }
\usepackage{subcaption}
\usepackage{wrapfig}

\usepackage{pifont}
\usepackage{diagbox}
\usepackage[normalem]{ulem} 

\usepackage{changepage}                 
\usepackage{lipsum}
\newlength{\offsetpage}
{\end{adjustwidth}}

\usepackage{bm}

\newcommand{\commentout}[1]{}
\newcommand{\iprod}[1]{\left \langle #1 \right \rangle}
\newcommand{\K}{\mathcal{K}}
\newcommand{\Hb}{\mathcal{H}}
\newcommand{\real}{\mathbb{R}}

\newcommand{\norm}[1]{\left\|#1\right\|}

\newcommand{\paren}[1]{\left(#1\right)}
\newcommand{\braces}[1]{\left\{#1\right\}}

\newcommand{\C}[1]{\Centerstack[c]{#1}}

\newcommand{\RA}[1]{\Centerstack[r]{#1}}
\newcolumntype{F}[1]{@{\hskip 1pt}>{\centering\arraybackslash}m{#1}@{\hskip 1pt}}

\definecolor{Gray}{gray}{0.9}
\newcolumntype{g}{>{\columncolor{Gray}}c}

\newcommand{\VLS}{
\centering
\setlength{\extrarowheight}{0pt}
\addtolength{\extrarowheight}{\aboverulesep}
\addtolength{\extrarowheight}{\belowrulesep}
\setlength{\aboverulesep}{0pt}
\setlength{\belowrulesep}{0pt}
}

\def\BibTeX{{\rm B\kern-.05em{\sc i\kern-.025em b}\kern-.08em
    T\kern-.1667em\lower.7ex\hbox{E}\kern-.125emX}}

\begin{document}

\title{An Efficient One-Class SVM for \\ Anomaly Detection in the Internet of Things}

\author{Kun Yang \\
Columbia University
\and  
{\bf Samory Kpotufe} \\
Columbia University
\and  
{\bf Nick Feamster} \\
University of Chicago
}

\date{}

\maketitle

\thispagestyle{plain}
\pagestyle{plain}

\begin{abstract}
    Insecure Internet of things (IoT) devices pose significant threats to
    critical infrastructure and the Internet at large; detecting anomalous
    behavior from these devices remains of critical importance, but fast,
    efficient, 
    accurate anomaly detection (also called ``novelty detection'') for these
    classes of devices remains elusive.
    One-Class Support Vector Machines (OCSVM) are one of the state-of-the-art
    approaches for novelty detection (or anomaly detection) in machine learning, due to their
    flexibility in fitting complex nonlinear boundaries between {normal} and
    {novel} data.  
    IoT devices in smart homes and cities
    and connected building infrastructure present a compelling use case for
    novelty detection with OCSVM due to the variety of devices, traffic
    patterns, and types of anomalies that can manifest in such environments.
    Much previous research has thus applied OCSVM to novelty detection for
    IoT.  Unfortunately, conventional OCSVMs introduce significant memory
    requirements and are computationally expensive at prediction time as the
    size of the train set grows, requiring space and time that scales with
    the number of training points. These memory and computational constraints
    can be prohibitive in practical, real-world deployments, where large training
    sets are typically needed to develop accurate models when fitting complex decision boundaries. In this
    work, we extend so-called Nystr\"om and (Gaussian) Sketching approaches to
    OCSVM, by combining these methods with clustering and Gaussian mixture
    models to achieve significant speedups in prediction time and space in
    various IoT settings, without sacrificing detection accuracy. 
\end{abstract}



\section{Introduction}

As devices from consumer electronics to building control systems increasingly
become connected to the Internet as part of the ``Internet of Things'' (IoT),
both this connected infrastructure and the network itself are subject to new
types of threats and vulnerabilities. A common approach to network defense
involves statistical anomaly detection, which aims to detect unusual activity
based on the observable properties of network traffic. 
One-Class Support Vector Machines (OCSVM) are one of the state-of-the-art
approaches for novelty detection in machine learning, due to their
flexibility, specifically their ability to identify a wide range of nonlinear
boundaries separating classes of data.  Such flexibility is naturally
appropriate
in scenarios with Internet of Things (IoT) devices and applications
applications, which naturally exhibit complexity due to the vast heterogeneity
of devices and the wide range of traffic patterns under various operating
modalities.  
It is therefore no surprise that OCSVMs have been been frequently
applied to novelty detection problems in IoT, with demonstrable efficacy at
detecting novel traffic patterns corresponding to either unseen modalities, or
to malicious activity~\cite{shilton2015dp1svm, lee2016packet,
mahdavinejad2018machine, al2020unsupervised,razzak2020randomized}. In the
context of security, novelty detection is often referred to as {\em anomaly
detection}; in this paper, we use the term {\em novelty detection} to refer to
the same class of algorithms, as the problem is equivalent. We prefer the use
of novelty detection in this paper because the classes of events that we aim
to detect include conventional anomalies (in the security sense), as well as a
broader class of novel events, activities, and devices that might be simply
``new'', though these new events may not necessarily have a negative
connotation.

The heterogeneity of IoT devices and operating regimes introduces a broad
class of activities (and corresponding network traffic patterns) that could be
classified as normal or novel. In contrast to general purpose computing
devices, where the main novel behavior of interest is typically a security
event such as an infection, networks with IoT devices may be concerned with a
far broader and more diverse set of anomalies, including physical device
failure, the introduction of rogue devices on the network, physical security
incidents, abnormal interactions with control systems, and so forth. The
devices themselves are also heterogeneous, with the normal operating regime for each
device type or manufacturer manifesting normal baseline patterns that are
distinct from one another. Such heterogeneity, both in terms of novelty and
device type, make OCSVMs an appealing tool for detecting novelty in these
contexts.

Many IoT deployments require fast novelty detection: in operational
deployments, there may be the need to quickly detect an attack, a rogue
device, or another malfunction. Unfortunately,
OCSVMs can be computationally expensive at \emph{detection time}, when
classifying new observations.  Given a new observation $x$ to classify as
normal or novel, detection consists of evaluating a scoring function
$f(x)$---of the form $\sum_{i =1}^n \alpha_i K(X_i, x)$, defined with respect
to training data $\braces{X_i}$ of size $n$ and a so-called \emph{kernel
function} $K$; such evaluation of $f(x)$ takes time and space $\Omega(n)$ for
typically large training data size $n$ in the thousands.  In the context of
IoT, each training datapoint $X_i$ represents a vectorized representation of
\emph{normal} traffic data over short time periods. Given an
Internet-connected device that is continuously generating network traffic,
detection using OCSVM is currently prohibitive in practice.

As such, the computational requirements of
OCSVMs can be prohibitive for practical deployments, where it is often
necessary to quickly detect anomalous events---if possible in field
deployments (e.g., embedded devices such as home network routers or embedded
sensors), where both computational and memory requirements may be limited.

{\bf Goals and Method.} The goal of this work is to speed up detection
time and reduce memory requirements of OCSVM, while preserving detection
performance, with specific applications in the IoT domain. Our
focus is on \emph{detection} time and space, as opposed to \emph{training}
time and space, as training can be done offline and is of less concern for a
practical deployment with space and time constraints.  
Novelty detection has long been an important problem area in network security,
although in many cases past work has explored novelty detection generally for
all types of network traffic. In this paper, we focus specifically on IoT
devices and activities because the traffic that these devices generate, and
the environments in which they operate, create a unique need and opportunity
for state-of-the-art novelty detection mechanisms such as OCSVM.  
First, because many IoT devices are task-specific, their modes of normal
operation can be characterized well. 

Consider, for example an Internet-connected appliance (e.g., a refrigerator);
the range of operating modes for such a device is far more limited than for a
general-purpose device such as a desktop computer, laptop, or smartphone,
which have a comparatively broader set of possible activities. Similarly, the
need for {\em efficient} detection, such as that we develop in this paper is
paramount, as fast attack detection is typically desirable, and in many IoT
devices involve the control over critical infrastructure (e.g., buildings,
industrial control systems), where rapid detection is essential. Finally, many
IoT devices operate in settings where memory and compute resources are
constrained; in such scenarios, the need for space-efficient detection
algorithms, such as the one we develop, are particularly
important.

Although novelty detection is an \emph{unsupervised problem}---
i.e., we only have access to \emph{normal data} as opposed to both normal and
novel datapoints---we draw initial inspiration from the related
\emph{supervised learning} method of Support-Vector-Machines (SVM), which,
similarly to OCSVM, uncovers linear relationships between classes of data.
Namely, various speedup approaches such as so-called Nystr\"om and Sketching
\cite{drineas2005nystrom, yang2017randomized} have recently been developed for
SVMs, which we aim to build on.  In our unsupervised IoT setting as we will
see, such speedup approaches require considerable adaptation if we hope to
preserve detection performance with respect to OCSVM. 

To better understand relevant discrepancies between unsupervised OCSVM and its
supervised counterpart, support vector machines (SVM) in applying speedup
methods, we need to get into a bit more detail. Most significantly, these
methods all operate on a so-called \emph{gram matrix} $\K \in \real^{n\times
n}$, encoding relations between datapoints, i.e., inner-products $\K_{i, j}
\doteq \phi(X_i)\cdot \phi(X_j)$ corresponding to an implicit data
transformation $x \mapsto \phi(x)$.  Operations on $\K$ are often the
bottleneck in training and prediction time, and approaches such as
\emph{Nystr\"om} and \emph{Sketching} portend to approximate $\K$ with a
lower-rank matrix $\K'$ that allows faster operations, while nearly
preserving the original relations between datapoints.  In particular, in the
case of SVM, the Nystr\"om or Sketching matrix $\K'$ manages to preserve the
same simple linear relationships between classes of datapoints encoded in
$\K$. In other words, one can simply proceed as usual with $\K'$ in place of
$\K$ and train a linear classifier.  Unfortunately as we will see (Section
\ref{sec:KJL}), in the unsupervised case of OCSVM, we lose the ability to
learn such simple linear relation between classes under
Nystr\"om or Sketching $\K'$, an issue particularly true in IoT, requiring a
different approach on top of Nystr\"om or Sketching. 

To address this issue of nonlinearity in the case of OCSVM after Nystr\"om or
Sketching, we rely on recent interpretations of these speedup approaches
\cite{yang2012nystrom, rudi2015less, kpotufe2020gaussian} whereas they might
be viewed as further data mapping $x\mapsto \phi(x) \mapsto \phi'(x)$ that
preserves distances between original transformed points $\phi(X_i),
\phi(X_j)$, even if linearity between classes is not preserved. As interpoint
distances are preserved, one might then expect that \emph{cluster} structures
are preserved, i.e., dense groups of points under $\phi$ remain clumped
together under $\phi'$. Building on this intuition, detection will therefore
just consist of flagging any future query point $x$ as \emph{abnormal} if
$\phi'(x)$ falls far from clusters in the remapped training data
$\braces{\phi'(X_i)}_{i=1}^n$.  To implement this idea, we model clusters in
$\braces{\phi'(X_i)}_{i=1}^n$ as components of a Gaussian Mixture Model (GMM),
which has the benefit of allowing for a simple detection rule based on
\emph{density levels} (see Section \ref{sec:detection}). Finally, as the GMM
model introduces a new hyperparameter on top of vanilla OCSVM, namely the
number $k$ of Gaussian components (or number of clusters), we further propose
a basic approach to automatically set such a parameter $k$ by estimating high
density regions of $\braces{\phi'(X_i)}_{i=1}^n$ via existing methods such as
\emph{QuickShift++} \cite{jiang2018quickshift++}. 

{\bf Results Overview.} We implement the above described approach, based on mapping the \emph{normal} training data as $\braces{\phi'(X_i)}_{i =1}^n$ using either Nystr\"om or a simple form of Sketching termed \emph{Kernel Johnson-Linderstrauss} (KJL) shown recently to preserve cluster structures w.r.t. to the original mapping $\phi$ induced by kernel methods such as OCSVM \cite{kpotufe2020gaussian}. For simplicity we will henceforth refer to these approaches respectively as OC-Nystr\"om and OC-KJL, where \emph{OC} stands for \emph{One Class} (as in OCSVM) to emphasize the unsupervised nature of these methods. We evaluate OC-Nystr\"om and OC-KJL, both with and without automatic GMM parameter selection, on multiple IoT datasets encoding a variety of detection use-cases of interest, e.g., detection of benign novelties such as traffic from new devices or new device modality, or detection of malicious activity from infected devices. 

To evaluate the effectiveness of our techniques in the context of IoT anomaly
detection, we evaluate our techniques on a variety of datasets---both public
network datasets that apply to IoT environments and datasets that we
have generated in the lab based on common interactions and scenarios with
consumer IoT devices. In addition to IoT-specific datasets, we also evaluate
our algorithms on several public datasets involving traffic generated by
general-purpose computing devices that are pertinent to IoT settings,
including distributed denial of service (DDoS) attack detection and the
appearance of novel device activity on the network.
\emph{The very nature of
these IoT use cases plays an important role towards achieving faster detection
time and space:} typical IoT devices, e.g., smart appliances, traffic
monitors, have few modalities of operations, inducing few \emph{clusters} of
normal traffic; as a result we can expect a small number $k$ of clusters,
i.e., GMM components needed to faithfully model normal operational traffic,
leading to smaller memory footprint and detection time complexity.  Our
results are as follows: 

\begin{itemize}[leftmargin=*]
    \item \emph{Significant reduction in detection time and space.} We observe typical detection time speedups (w.r.t. the baseline OCSVM) 
    in factors between 14 to 20 times using either of OC-Nystr\"om or OC-KJL, and reaching up to 40+ times for some datasets. Typical space complexities decrease by factors of 20 or more w.r.t. OCSVM. 
    \item \emph{Equivalent or improved detection performance.} 
    Given that detection performance of any machine learning method depend
        crucially on hyperparameter choices, we consider two situations: (1)~where hyperparameters are adequately calibrated using side data (i.e.,
        a small validation set independent of future test data), and (2)~a
        situation where such side data might be missing and basic
        rules-of-thumb are employed to select hyperparameters.  Such a
        situation might arise in IoT settings where some activities and
        devices might be labeled, but the vast majority remain unlabeled due
        to the large scale and heterogeneity of datasets.
\end{itemize}
    
Upon proper calibration of all three procedures, both OC-Nystr\"om or OC-KJL
achieve detection performance on par with the baseline OCSVM as measured by
Area-Under-the-Curve (AUC). In fact, both slightly outperform OCSVM in some
cases, which is likely due to the fact that the new mapping $\phi'$ of the
data acts as a lower-dimensional projection which at times recovers intrinsic
structure not present in abnormal traffic. 
    
In the second situation, i.e., under rules-of-thumb choices of the main
hyperparameter shared by all three procedures, i.e., a so-called \emph{kernel
bandwidth} parameter, OC-Nystr\"om and OC-KJL (with automatic choices of
number of GMM components $k$) attain at least 0.85\% of OCSVM's AUC on most
datasets, and even manages significant improvements in AUC over that of OCSVM
on many datasets.  Given the lack of proper calibration however, we observe
some rare situations where AUC degrades more considerably w.r.t. that of
OCSVM.  These are included to give a fair and broad sense of the range of
performance one could potentially observe in practice.  

\section{Related Work}

\subsection{Network anomaly detection} Anomaly detection in networks is a
widely studied problem; a wide range of techniques have been applied to this
problem over the past several decades. Ahmed et al. provide a more complete
survey of these techniques~\cite{ahmed2016survey}; we briefly overview some of
the general classes of techniques.  Various supervised learning techniques
have been applied to the problem of network anomaly detection, including
support vector machines~\cite{eskin2002geometric}, Bayesian
networks~\cite{kruegel2003anomaly}, sequential hypothesis
testing~\cite{jung2004fast},
and neural networks~\cite{hawkins2002outlier,wang2017hast}. In many of these
cases, supervised learning has been applied in a very specific context, such
as detecting port scans~\cite{jung2004fast} or web-based
attacks~\cite{kruegel2003anomaly}, where obtaining a labeled dataset for the
specific attack or anomaly of interest is feasible.

In contrast to some of the previous work on supervised anomaly detection,
which have typically involved the detection of a {\em specific} type of
attack, general supervised learning approaches for general anomaly detection
on network traffic are often impractical, particularly in IoT settings, where
it may be difficult to label or characterize a complete set of anomalies,
given the particularly large and diverse set of IoT devices and set of
possible activities. In the case of IoT, large labeled datasets of devices and
activities do not exist; furthermore, due to the diverse nature of IoT devices
and modes of interaction, anomalies may differ significantly across types of
devices, environments, and modes of interaction, and thus a labeled dataset in
one scenario is unlikely to transfer to other environments.

Common unsupervised approaches have involved the use of principal component
analysis~\cite{shyu2003novel,lakhina2004diagnosing} and generalized likelihood
ratio~\cite{thottan2003anomaly}. Anomalies in these settings encompass events that
include network failures, large-scale shifts in traffic, performance problems,
and denial of service (DoS) attacks. These works focused largely on the
detection of anomalous events in the context of traffic flows that traverse
wide-area backbone networks and are generally concerned with abrupt shifts in
traffic volumes that are visible in aggregated traffic statistics. The
techniques have also been implemented on offline traces without particular
attention to time or space efficiency. Principal component analysis in
particular has proved problematic in the context of network anomaly detection
due to the fact that transforming network traffic into a matrix representing a
multidimensional timeseries involves quantization and discretization that
render the resulting underlying models brittle~\cite{ringberg2007sensitivity}.
In particular, Ringberg et al. found that when applying PCA to network traffic
anomaly detection, the false positive rate is sensitive to the selection of
the number of principal components in the normal subspace and the level of
traffic aggregation~\cite{ringberg2007sensitivity}.

\subsection{Anomaly detection in IoT} Over the past several years,
unsupervised learning techniques have been developed for novelty detection
specifically for IoT devices and activities; one-class SVM has been
particularly effective for detecting anomalies in IoT
settings~\cite{shilton2015dp1svm, lee2016packet, mahdavinejad2018machine,
al2020unsupervised,razzak2020randomized}. OCSVM is particularly appropriate
for novelty detection in IoT due to its ability to learn complex, non-linear
decision boundaries, which can be particularly important in IoT environments
where activities are diverse and heterogeneous. Unfortunately, however,
despite its efficacy in these settings, OCSVM can be particularly costly in
terms of both time and memory requirements, rendering the previous work
impractical for many deployment settings where novelty detection algorithms
would be deployed in practice. Specifically, IoT deployments involve the
deployment of resource constrained devices; in the case of consumer IoT
deployments, for example, anomaly detection systems may need to operate on
home routers, where processing and memory capacity is limited.  The algorithms
we develop in this paper achieve speedup of up to 40 times as compared to the
best-known implementations of OCSVM, thus making it possible to deploy these
anomaly detection algorithms in practice in IoT settings.  To demonstrate this
feasibility, we evaluate the real-time performance and memory requirements of
our algorithms on embedded single-board computers that are often deployed
in home network settings.


\section{Background on Methods} 

\subsection{(Gaussian Kernel) OCSVM}\label{sec:OCSVM}
\begin{figure*}[!htbp]
    \centering
    \includegraphics[width=0.85\textwidth]{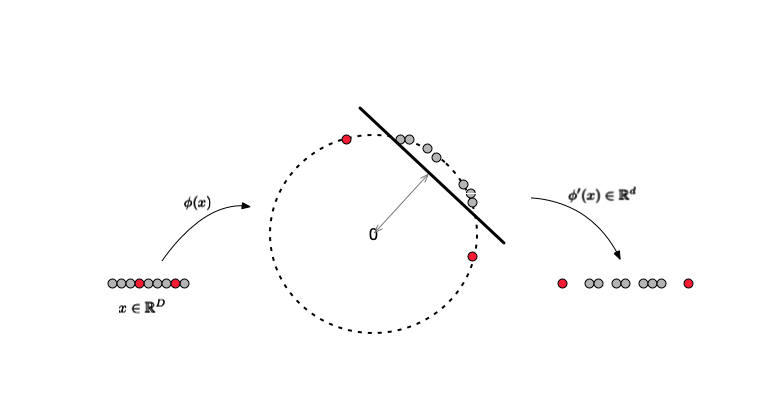}
    \caption{OCSVM maps datapoints $x\in \real^D$ as $\phi(x)$ in infinite-dimensional space, inducing linear separation between classes of points (red and gray datapoints). Since only the \emph{normal} data (gray points) is available at training time, finding a hyperplane separating normal and abnormal data might seem a priori unfeasible. Fortunately $\phi$ maps all datapoints to the surface of an infinite-dimensional sphere, so a separating hyperplane can be found using normal data alone, which separates such data as much as possible from the center of the sphere, i.e., from the 0 vector. Such linear separability from $0$ is often lost by Nystr\"om and Sketching as the resulting embedding $\phi'$ no longer maps to the surface of a sphere; however, cluster structures evident under $\phi$ are maintained by $\phi'$.}
    \label{fig:basicOCSVM}
\end{figure*}

{\bf Basic background.} OCSVM first maps data $x\in \real^D$ as $\phi(x)$ into an infinite dimensional space $\Hb$ (a so called \emph{reproducing kernel hilbert space} or RKHS). As a hilbert space, $\Hb$ admits basic vector operations as in Euclidean $\real^D$, in that it has a well-defined inner-product 
$\iprod{\phi(x), \phi(x')}$ inducing a norm $\norm{\phi(x)}^2 = \iprod{\phi(x), \phi(x)}$ and hence a notion of distance between points and space geometry (clusters, linear projections, hyperplanes, spheres, etc). All that is therefore needed for geometric operations is access to the inner-product operation $\iprod{\cdot, \cdot}$, which is readily provided by RKHS theory: for any datapoints $x, x' \in \real^D$, there exists a so-called \emph{kernel function} $K$ satisfying $K(x, x') = \iprod{\phi(x), \phi(x')}$. Therefore, given access to $K$, the mapping $\phi$ need not be explicitly computed, as all geometric operations are implicit through $K$ alone, and in particular, all geometric operations involved in learning a hyperplane separating classes of points are thus determined by $K$ alone. The most common kernel function in machine learning, and especially in OCSVM, is the \emph{Gaussian} kernel 
$K(x, x') = C\cdot \exp\paren{- \norm{x- x'}^2/2h^2}$ (for a \emph{bandwidth} hyperparameter $h$ to be chosen in practice, and a normalizing constant $C = C(h)$).


{\bf Key intuition and operations.} A main intuition behind the mapping $\phi$, implicit in both supervised SVM and unnsupervised OCSVM, is that it manages to \emph{separate} classes of data, i.e., pull corresponding datapoints far apart in $\Hb$, even when they are not easily separable in their original representation in $\real^D$.
This is illustrated in Figure \ref{fig:basicOCSVM}. It follows that, after the mapping $\phi$, the data might become linearly separable in $\Hb$, i.e., the two classes of data, \emph{normal} and \emph{abnormal}, fall on different sides of a hyperplane in $\Hb$. Therefore, in supervised learning (e.g. with SVM) where we have access to both classes of data at training time, we simply would learn a hyperplane that most faithfully separates the training data into the two class labels. However, in the case of OCSVM, only one class is available during training, namely \emph{normal} data. It is therefore unclear how to separate it from unseen \emph{anomaly} data. The main insight is in that, if the kernel $K$ satisfies $K(x, x) = C$ for some constant $C$, as with the Gaussian kernel, then all points $x\in \real^D$ are mapped in $\Hb$ to the surface of a sphere of radius $\sqrt{C}$, since $K(x, x) = \norm{\phi(x)}^2 = C$. It follows that if the two classes are linearly separable, then they can be separated by a hyperplane that puts maximal margin between the \emph{normal} class and the center of the sphere, since unseen anomaly data is also constrained to map to the surface of the sphere. This is illustrated in Figure \ref{fig:basicOCSVM}. 

\begin{wrapfigure}{r}{.5\textwidth}
\centering
    \includegraphics[width=0.45\textwidth]{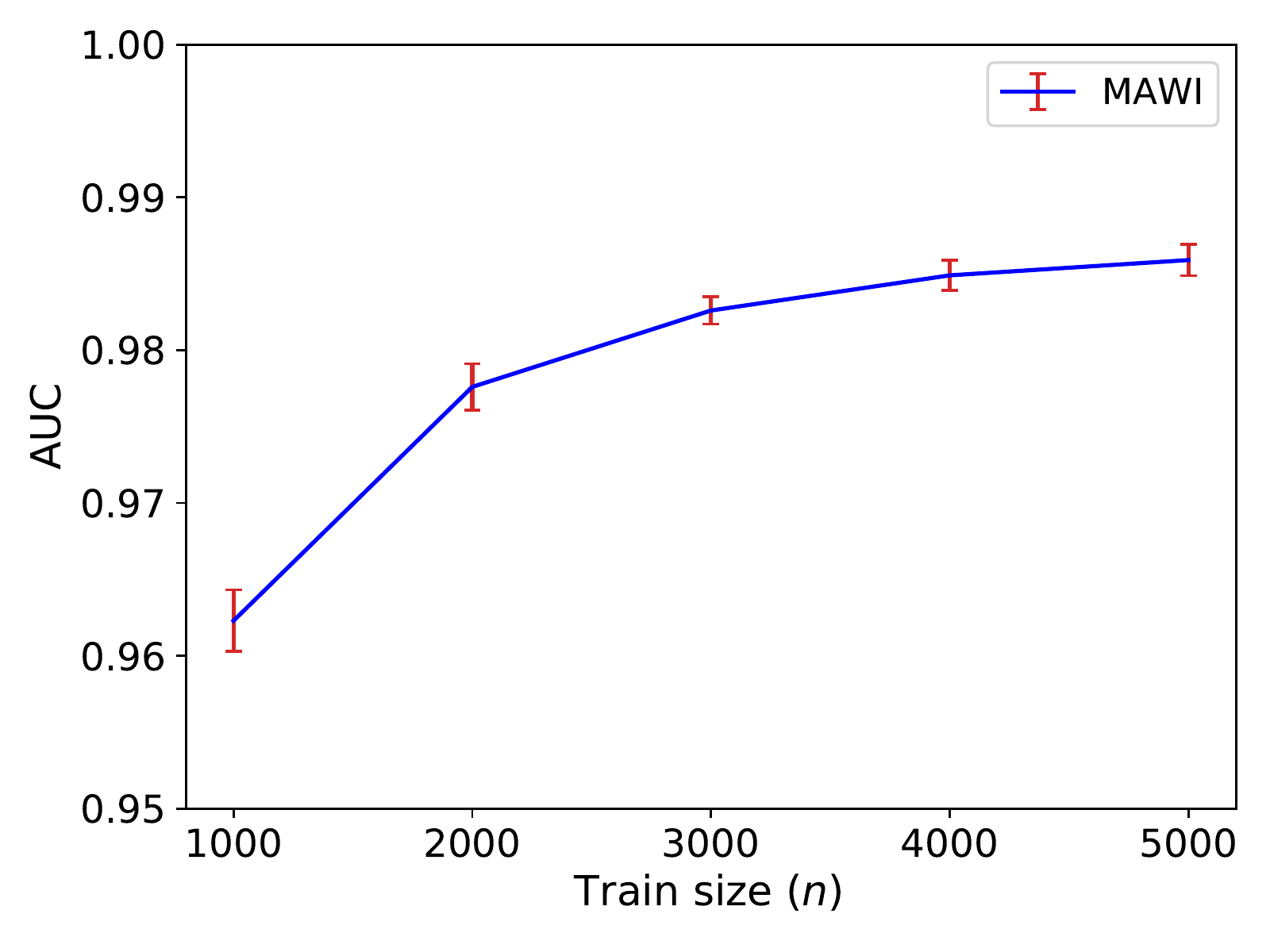}
\caption{Detection performance of OCSVM, as captured by AUC, increases with training size $n$. Unfortunately, so do the detection time and space complexity of OCSVM.}
\label{fig:speedup} 
\end{wrapfigure}

OCSVM thus, using normal data $\braces{X_i}_{i=1}^n$ alone, returns a hyperplane that isolates normal data from future anomalous observations. Such a hyperplane can be estimated without actually computing $\phi(X_i)\in \Hb$, simply through geometrical operations encoded by all pairwise inner-products $\iprod{\phi(X_i), \phi(X_j)}$ given by $K(X_i, X_j)$. These inner-products are encoded for convenience in a so-called \emph{gram matrix} $\K\in \real^{n \times n}, \ \K_{i, j} = K(X_i, X_j)$ so the training phase just operates on $\K$ to return an implicit representation of the separating hyperplane in the form of coefficients 
$\braces{\alpha_i}_{i =1}^n$ and a threshold $\alpha_0$ used as follows: 
\begin{quote}
    A future test point $x\in \real^D$, is deemed anomalous if it maps as $\phi(x)$ to \emph{the wrong side} of the hyperplane, that is, if $f(x) \doteq \sum_{i=1}^n \alpha_i K(X_i, x) < \alpha_0$. 
\end{quote}

In other words, as in Euclidean spaces, $f(x)$ can be viewed a the projection of $\phi(x)$ onto a vector normal to the separating hyperplane, and the $\alpha_i$'s are coefficients determining this vector.

{\bf Detection time and space.} It should be clear by now that computational complexity is determined by the number $\tilde n\leq n$ of nonzero $\alpha_i$'s. The corresponding datapoints $X_i$'s are called the \emph{support vectors}, and have to be kept in memory to estimate $f(x)$. Thus the OCSVM detector takes space $\tilde n \cdot (D+1)$, while computation time for $f$ is $\Omega (\tilde n \times D)$. Unfortunately, it is often the case that $\tilde n = n$ or is of the same order, while the larger $n$, the more accurate the detector is (Figure \ref{fig:speedup}). 

\subsection{Nystr\"om and KJL sketching}\label{sec:KJL}

A main approach adopted recently to speedup training time, e.g., in the context of SVMs, is to reduce operations on the gram matrix $\K\in \real^{n\times n}$ by approximating it with a rank $d\ll n$ matrix $\K'\in \real^{n\times n}$ that might induce faster operations, while preserving much of the geometry induced by the kernel $K$ on the implicit mapping $\braces{\phi(X_i)}_{i =1}^n \in \Hb$. These come in different forms under the name of Nystr\"om and Sketching. In particular, in some implementations, we can view $\K'$ as inducing a new mapping $x\mapsto \phi'(x)$ for $\phi'(x) \in \real^d$, i.e., a low-dimensional mapping that preserves some geometry in $\Hb$. 

Critically, as explained in the introduction, such $\phi'$ often no longer allows for linear separability from $0$ -- i.e., using just one class in the training data -- as in the case of the original OCSVM map $\phi$, since the remapped data $\braces{\phi'(X_i)}_{i =1}^d$ no longer lies on the surface of a sphere (see Figure \ref{fig:basicOCSVM}). However, cluster structures uncovered by the original $\phi$ are preserved, since $\phi'$ preserves interpoint distances (see e.g. \cite{calandriello2018statistical, kpotufe2020gaussian}), which we build on in Section \ref{sec:detection} below. 

{\bf Embedding $\phi'$.} Crucially, in order to leverage cluster structures towards efficient outlier detection, we make the embedding $\phi'$ explicit -- as opposed to operating on $\K'$ -- and work directly in $\real^d$. This is based on recent reinterpretations of forms of Nystr\"om and Sketching as low-dimensional projections \cite{yang2012nystrom, kpotufe2020gaussian}. In both cases, let 
$S_m$ denote a random subsample of size $m \ll n$ of the training data $S_n \doteq \braces{X_i}_{i=1}^n$ (w.l.o.g., we can let 
$S_m \doteq \braces{X_i}_{i=1}^m$). Furthermore, for any subset of indices $I, J \subset \braces{1, \ldots, n}$, let $\K_{I, J}$ denote the submatrix of $\K$ corresponding to rows in $I$, and columns in $J$. Then, for $I = \braces{1:m}$ and $J = \braces{1: n}$, we will consider the submatrices, 
$\K_{I, I}\in \real^{m\times m}$ -- i.e., the gram matrix on $S_m$, and $\K_{I, J}\in \real^{m\times n}$, the gram submatrix of inner-products between $S_m$ and $S_n$.  
\begin{itemize}[leftmargin=*]
    \item \emph{Nystr\"om.} Let $K_{I, I}^{-1}$ denote a rank $d$ pseudo-inverse of $K_{I, I}$; then setting 
    $\K' = K_{I, J}^\top \cdot K_{I, I}^{-1} \cdot K_{I, J}$, the problem is to come up with $\phi'\in \real^d$ such that $\iprod{\phi'(X_i),\phi'(X_j)}$ is exactly $\K'_{i, j}$. Recalling a bit of linear algebra, we can see that a suitable $\phi'$ can be defined as follows \cite{yang2012nystrom}. Let $\Lambda\in \real^{d\times d}$ denote the diagonal matrix containing the top $d$ eigenvalues $\lambda_1, \ldots ,\lambda_d$ of $\K_{I, I}$, and $V = [v_1,\ldots, v_d]\in \real^{m\times d}$ contains the corresponding (column) eigenvectors $v_i$'s. Now, for any $x\in \real^D$, let $K(x)$ denote the vector $[K(x, X_1), \ldots, K(x, X_m)]^\top$, we then have 
    \begin{equation}
        \phi'(x) \doteq  P\cdot K(x), \textit{ where we let }  P\doteq \Lambda^{-1/2}\cdot  V^\top. 
        \label{eq:nystrom}
    \end{equation}
    We can verify that setting $K_{I, I}^{-1} = V \cdot \Lambda^{-1} \cdot V^\top$, indeed recovers $\K'$ as defined above.
    
    \item \emph{KJL Sketching.} In general, Sketching consists of multiplying a gram matrix $\K$ (or $K_{I, I}$) by a matrix $Z$ with random entries. It was recently shown \cite{kpotufe2020gaussian} that when $Z$ has i.i.d. $\mathcal{N}(0, 1)$ Gaussian entries, sketching can be understood as a random projection operation in $\Hb$, leading to the following mapping $\phi' \in \real^d$. For any $x \in \real^D$, let $K(x)$ again denote the vector $[K(x, X_1), \ldots, K(x, X_m)]^\top$, and let $Z\in \real^{d\times m}$ with random $\mathcal{N}(0, 1)$ entries. We then have: 
    \begin{equation}
        \phi'(x) \doteq P\cdot K(x) \textit{ where we let } P \doteq Z\cdot K_{I, I}. \label{eq:kjl}
    \end{equation}

\end{itemize}

{\bf Embedding time and space.} Notice that in both cases of Nystr\"om and KJL, we only have to retain $P\in \real^{d\times m}$ at testing time, along with the $m$ datapoints in $S_m$. In other words, this contributes a space complexity of exactly $m\cdot (d+ D)$. Similary the time complexity of computing $\phi'(x)$ just depends on these 3 parameters $m, d, D$ but not on the training size $n$. 

\begin{figure}[!htbp] 
\centering
    \includegraphics[width=0.85\textwidth]{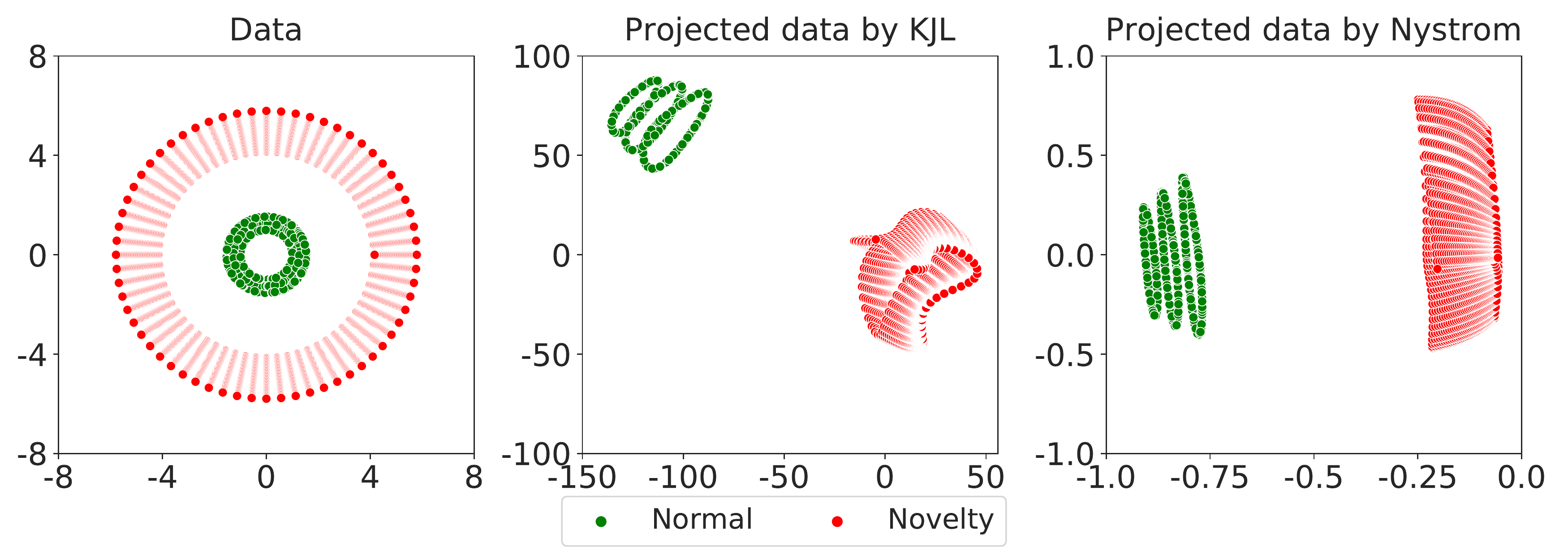}
 \caption{\small Clusters after mapping $\phi'$: the simulation data \emph{Cluster in Cluster} has 5000 points, shown before and after KJL/Nystr{\"o}m mapping. The KJL/Nystr{\"o}m mapping $\phi'$ -- shown on the right' retains the clusters uncovered by the initial kernel mapping $\phi$. }
\label{fig:projection} 
\end{figure}

As it turns out $m, d$ can be kept considerably smaller than $n$, while achieving the benefits of both methods. This is illustrated in Figure \ref{fig:projection}, on simulated data of size $n = 5000$, with two classes that are not easily clustered in $\real^D$, but which are clusterable not only in $\Hb$, but also after Nystr\"om of KJL. In that simulation we used $d=2$, and $n = 200$. Similar settings are used for our experiments on real-world IoT data (see experimental setup in Section \ref{sec:experimentalsetup}).


\section{Efficient Detection Procedures}\label{sec:detection}
\begin{wrapfigure}{l}{.45\textwidth}
\centering 
    \includegraphics[width=0.45\textwidth]{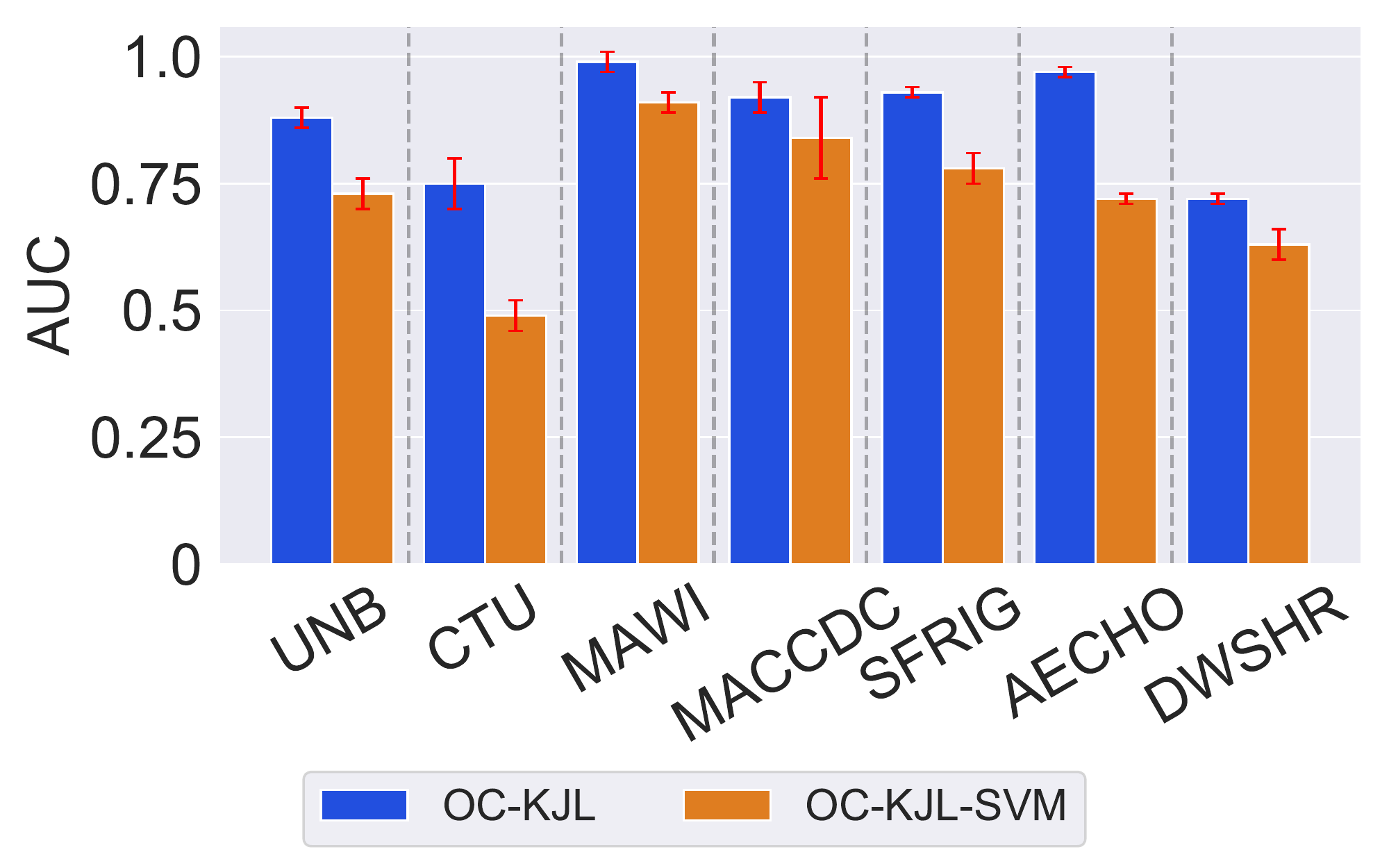}
\caption{OC-KJL vs. OC-KJL-SVM.}
\label{fig:iat_size-best-oc_kjl_vs_oc_kjl_svm-full}
\end{wrapfigure}

Once the data is mapped to $\real^d$ as $\braces{\phi'(X_i)}_{i =1}^n$ through Nystr\"om or KJL, our next step is to learn an efficient model of the normal class embedded in $\real^d$. Recall that cluster structures are preserved, but not necessarily linear separability from $0$ (see e.g. simulation of Figure \ref{fig:projection} where the normal class is not necessarily linearily separable from the origin $0\in \real^2$). Looking somewhat ahead, this intuition is validated with the results of Figure \ref{fig:iat_size-best-oc_kjl_vs_oc_kjl_svm-full} where we compare fitting a linear separator after KJL projection (denoted OC-KJL-SVM) to our proposed method (OC-KJL) soon to be described. The detection performance metric is the Area-Under-the-Curve (AUC) -- described in detail in Section \ref{par:metrics} -- which is consistently higher for OC-KJL across datasets. 

\begin{figure}[!hbp]
\centering
\includegraphics[width=0.5\textwidth]{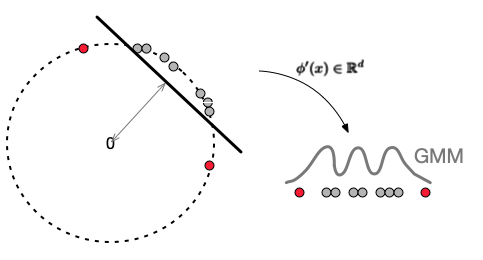}
  \caption{GMM on remapped \emph{normal data}}
  \label{fig:remappedGMM}
\end{figure}

A natural idea therefore is to flag future points as novelty if they fall far from clusters in $\braces{\phi'(X_i)}_{i =1}^n$. Suppose there are $k$ clusters, then 
a simple implementation of this idea is to fit a \emph{Gaussian Mixture Model} (GMM) to the normal remapped data $\braces{\phi'(X_i)}_{i =1}^n$, with $k$ components that  encode clusters within; such a GMM is obtained as a probability density of the form 
\begin{equation}
    f(z) = \sum_{l=1}^k \pi_l \cdot \mathcal{N}\paren{z; \mu_l, \Sigma_l}, \, \textit{ for any } z\in \real^d,
\end{equation}
where $\mathcal{N}\paren{z; \mu_l, \Sigma_l}$ denotes a Gaussian density with mean $\mu_l$ and covariance $\Sigma_l$ evaluated at $z$, and $\pi_l$'s denote the probablity or mass of each cluster $l \in \braces{1, \ldots, k}$ and sum up to 1. Such a density $f$ would have \emph{modes} $\mu_l$, i.e., local maxima a.k.a. \emph{high-density cores}, centered on clusters, as illustrated in Figure~\ref{fig:remappedGMM}. 

Once $f$ is learned, detection simply consists of flagging $x$ as an novelty if $f(\phi'(x))$ is small than a threshold $t$. In practice such a threshold can be picked depending on the amount of tolerable false positive; for instance if we want at most $5\%$ false positives, we might set $t$ as the 95th quantile of $f$ values (in decreasing $f$ order) on the negative data, i.e., on the embedded normal data $\braces{\phi'(X_i)}_{i=1}^n$. In our experiments below we will report the performance of detectors across all such thresholds choices, as captured by AUC (see Sections \ref{sec:experimentalsetup} and \ref{sec:results}).

{\bf Choice of number of components $k$.} As discussed earlier, we may automatically chose the number of components $k$ by first identifying the number of \emph{high density regions} in the mapped data $\braces{\phi'(X_i)}$. This might be done a number of ways, and we propose to use available density-mode estimators such as from the \emph{Meanshift} family \cite{comaniciu1999mean}; these are procedures that automatically identify the modes, i.e. local maxima, of the underlying data density, which in simple terms are just the regions of highest density in the data. In particular, in this work we employ a recent fast version of these mode estimators denoted \emph{QuickShift++} \cite{jiang2018quickshift++}, which automatically returns points in locally high density regions of the data, \emph{with no a priori knowledge of the number of such regions}, which we will identify with clusters. However if labeled side data is available to cross-validate for the OCSVM, or Nystr\"om and KJL bandwidth parameter $h$, the same data can be used to choose $k$ (see Section \ref{sec:experimentalsetup}).

{\bf Meta Procedures.} The resulting OC-Nystr\"om and OC-KJL approaches are summarized below. Given a Gaussian kernel $K$ with bandwidth $h$, embedding choices $m, d \ll \textit{training size } n$: 

\begin{tcolorbox}[colframe=gray,boxsep=3pt,left=3pt,right=3pt,top=3pt,bottom=3pt]\small 
\underline{\bf Training:} Given normal data $\braces{X_i}_{i = 1}^n \in \real^D$ do: \\ 
\vspace{-6pt}

- Embed data as $\braces{\phi'(X_i)}_{i = 1}^n \in \real^d$ via Nystr\"om \eqref{eq:nystrom} or KJL \eqref{eq:kjl} \\
\vspace{-6pt}

- Parameter $k$ is passed in, or chosen via Quickshift++ on \\
$\quad$ embedded data $\braces{\phi'(X_i)}_{i = 1}^n \in \real^d$\\
\vspace{-6pt}

- Estimate a GMM density $f$ with $k$ components on $\braces{\phi'(X_i)}_{i = 1}^n$ \\
\vspace{-6pt}

- {\bf Return} GMM $f$, along with projection $\phi'$ (i.e., matrix $P$ and subsample $S_m$) $\qed$ \\
\hrule
\vspace{-6pt}
 $\,$ \\ $\,$ \\
\underline{\bf Detection:} Given new $x \in \real^D$, and model $\paren{\phi', f}$, do: \\
\vspace{-6pt}

- Embed $x$ as $\phi'(x)$ into $\real^d$ \\
\vspace{-6pt}

- Flag $x$ as novelty iff $f(\phi'(x)) \leq \textit{ threshold } t$ $\qed$
\end{tcolorbox}
{\bf Detection time and space.} As described in Section 2.2., saving $\phi'$ takes space $m\cdot(d +D)$, while $f$ now takes additional space $k\cdot(d + d^2)$ for GMM parameters. As $m, d$ can be chosen small, detection time mostly depends on $k$; fortunately, as discussed in the introduction, $k$ can be chosen small (between 1 and 20 in our experiments) as clusters naturally correspond to the typically few modes of normal operation of IoT devices. 

{\bf OC-Nystr\"om vs OC-KJL.} As we will see in the results Section \ref{sec:results}, both procedures achieve our intended goal of efficiency while maintaining detection performance on par with that of OCSVM; while advantages vary across datasets, OC-Nystr\"om tends to trade a bit of efficiency for better detection, as its embedding might require larger $k$ values at time.


\section{Experimental Setup} \label{sec:experimentalsetup}


\subsection{Datasets}

We consider a combination of publicly available traffic traces and traces
collected on private consumer IoT devices. We aim to evaluate representative
set of devices, from multi-purpose devices such as laptop PCs, and Google
Home, to less complex electronics and appliances with few modes of operations
such as smart cameras or smart fridges. Furthermore we aim at a 
representative set of \emph{novelties}, from benign novelties (new activity,
or a new device type), to novelties due to malicious activities (DDoS attack).
Table \ref{dataset_sources} describes these datasets, and types of novelty
being detected. 

Interestingly, while some of the devices such as PC's are multipurpose and as such might display a significant number of modalities, including them allows us to test how well our approach scales. In particular, we will see that efficient detection is possible even in such cases, as even then $k \leq 20$ clusters suffice to maintain detection performance over these datasets, while we keep $d=5$, and $m = 100$, i.e., uniformly low.

\begin{table}[htbp!]
\centering
\VLS
\caption{Datasets.}
\begin{tabular}{|F{2.5cm}|m{8cm}|F{1.8cm}|F{1.6cm}|F{1.9cm}|} 
\toprule 
\C{Reference} & \multicolumn{1}{c|}{Description} & \C{Devices} & \C{Type of\\ Novelty}  &\C{Dimension $D$\\ (IAT+SIZE)}  \\
\hline
\C{Lab IoT SFRIG} & {Data traces are generated by a Samsung Fridge (SCam) with IP '192.168.202.43' in a private lab environment. It has two types of traffic traces labeled as \emph{normal} when there is no human interaction, and \emph{novel} when being operated by a human (such as, open the fridge).} 
  &   One fridge  & \C{Novel\\ activity}  & 23\\
\midrule
\C{Lab IoT AECHO} & {Data traces are generated by a Amazon ECHO (AECHO) with IP '192.168.202.74' in a private lab environment. It has two types of traffic traces labeled as \emph{normal} when there is no human interaction, and \emph{novel} when being operated by a human (such as, buy food by the AECHO). }
  &   One Amazon ECHO  & \C{Novel\\ activity}  &51 \\
 \midrule
\C{Lab IoT DWSHR} & {Data traces are generated by a dishwasher (DWSHR) with IP '192.168.202.76' in a private lab environment. It has two types of traffic traces labeled as \emph{normal} when there is no human interaction, and \emph{novel} when being operated by a human (such as, open the dishwasher). We also add another novel traffic (such as, open a washing machine) collected from a washing machine (with IP  '192.168.202.100') into \emph{novel} to get a bigger testing set. }
  &   One dishwasher and one washing machine  & \C{Novel\\ activity}   & 21\\
 \midrule
CTU IoT \cite{ctu2019iot}  &
 {Bitcoin-Mining and Botnet traffic traces generated by two Raspberries; we
    use Botnet traffic (with IP '192.168.1.196') as \emph{normal},
    Bitcoin-Mining traffic (with IP '192.168.1.195') as \emph{novel}.} &
    \C{Two \\ infected\\Raspberry\\Pis}  & \C{Novel\\ (infected)\\ device}  & 23 \\ 
\hline
\hline
UNB IDS\cite{cicids2017}  & {Normal traces are generated by one personal computer (PC) with IP address is '192.168.10.9'. Attack traces are generated by three PCs, with IP addresses are '192.168.10.9', '192.168.10.14', and '192.168.10.15'.} & 
Four PCs & DDoS attack & 47 \\
\hline
MAWI\cite{mawi2019normal} & {Normal traffic are collected on July 01, 2020; we choose one kind of traffic generated by a PC with IP '203.78.7.165' as \emph{normal}, and another kind of traffic generated by a PC with IP address '185.8.54.240' as \emph{novel}.}  &  Two PCs   &   \C{Novel\\ (normal)\\ device}  & 121\\ 
\hline
\C{MACCDC\cite{maccdc2012normal}}& {Data traces are collected in 2012. We choose one kind of traffic generated by a PC with IP '192.168.202.79' as \emph{normal} and one kind of traffic generated by a PC with IP '192.168.202.76' from another pcap as \emph{novel}.}
  &  Two PCs  & \C{Novel\\ (normal)\\ device}  & 25\\
\midrule 

\bottomrule
\end{tabular}

\label{dataset_sources}
\end{table}

\subsection{Data Representation}\label{sec:features}
Our unit of measurement consists of traffic flows, described below, i.e., as we aim to flag flows as normal or novel. 
\paragraph{Obtaining flows}
We parse bidirectional flows from datasets in Tab. \ref{dataset_sources} using
Scapy \cite{scapy} and extract interarrival times and packet sizes as our features.  Given that certain devices can have arbitrarily long
flows, we truncate each flow from a given dataset to have duration at most
that of the 90th upper-percentile of flow durations in the dataset.
Henceforth, a \emph{flow} refers to these choices of flows involving
truncation. Information on resulting data sizes, i.e., number of obtained normal and novel flows, are given in Table \ref{tab:dataset_details} of the Appendix. These are then subsampled from to obtain random instances of training of size $n = 5000$, validation size 150, and test data size 600. 

\paragraph{Extracted Features}
Every flow is represented as a vector of the inter-arrival times between packets, i.e., {in microseconds} elapsed between consecutive packets, along with the size {in bytes} of each packet in the flow (IAT+SIZE). 

We make this choice of features as it results in competitive detection accuracy for OCSVM, as compared with other popular choices. This is demonstrated, e.g., against 2 common  alternative feature choices, namely STATS, and SAMP\_SIZE, as shown in the Table \ref{tab:different_feature} below. 

\begin{table}[H]
\centering
\VLS
\caption{Average AUCs accross alternative choices of features.}
\label{tab:different_feature}
\begin{tabular}{|F{2.8cm}|c|c|c|c|c|c|c|}
\toprule
Dataset      & CTU          & MAWI    & SFRIG   & DWSHR \\
\midrule
IAT+SIZE  & 0.65$\pm$0.01 & 0.99$\pm$0.00 & 0.93$\pm$0.00 & 0.71$\pm$0.01\\
\midrule
STATS+HEADER    & 0.60$\pm$0.01 & 1.00$\pm$0.00 & 0.92$\pm$0.01  & 0.64$\pm$0.00\\
\midrule
SAMP\_SIZE  & 0.61$\pm$0.01 & 0.98$\pm$0.00 & 0.93$\pm$0.01 & 0.70$\pm$0.00\\ 
\bottomrule
\end{tabular}
\end{table}

STATS+HEADER corresponds to common statistics on flows, e.g., flow duration,
mean, standard deviation and quantiles of packet sizes, in addition to packet
header information \cite{yang2020comparative} as described in detail in the
Appendix.


In Appendix \ref{appendix_stats+header}, additional we show results for
experiments using the alternative features set, STATS+HEADER, to
demonstrate that significant savings in time and space over baseline OCSVM, 
do not depend on any particular choice of data representation. This result is
expected because the main source
of savings in both time and space results from our succinct finite-dimensional modeling of the
infinite-dimensional representation inherent in OCSVM.



\subsection{Implementation Details and Hyperparameter Choices}
\label{sec:hyperparamTuning} All detection procedures are implemented in
Python, calling on the \texttt{scikit-learn} package for existing procedures
such as OCSVM and GMM.  While OCSVM training uses the standard \texttt{libsvm}
package, we re-implemented its detection routines (as described in Section
\ref{sec:OCSVM}) using \texttt{numpy} to ensure fair, apples-to-apples
execution time comparison with OC-Nystr\"om and OC-KJL, which are implemented
in \texttt{numpy}, a Python library which calls on fast algebraic operations
and parallel processing on multicore machines \cite{harris2020array}.  The
Nystr\"om and KJL projections are implemented as described above, and we plan
to release the code on Github~\cite{ockjl}. 

\paragraph{Two Training Scenarios} As discussed in the introduction, we
consider two main practical scenarios: one where some small
amount of labeled \emph{novelty} data is available to \emph{validate}
hyperparameter choice, as in a controlled lab environment, and one with
no such labeled validation data, where we have to result to default choices of
hyperparameters. We note here, that while each detection procedure may have
many internal parameters, this distinction in scenarios only applies to two
key choices of hyperparameters: 

\begin{itemize} [leftmargin=*]
\item {\it Kernel Bandwidth $h$.} For all methods, i.e., OCSVM, OC-Nystr\"om, OC-KJL, we use a Gaussian kernel of the form $K(x,x') \propto \exp(-\|x-x'\|^2/h^2)$, where the \emph{bandwidth} $h$ is to be picked as a quantile of ${n \choose 2}$ distances between the $n$ training datapoints. In all our results we consider 10 quantiles $[0.1, 0.2, \ldots, 0.9] \cup \{0.95\}$ of increasing interpoint distances. 

\item {\it Number of GMM components $k$.} As explained above, OC-Nystr\"om and OC-KJL also require a choice of number of GMM components to fit. We consider choices in the range $[1, 4, 6, 8, 10, 12, 14, 16, 18, 20]$. Thus the number of components, or \emph{clusters} $k$ is capped at 20, as the devices being monitored are expected to display relatively few modes of operations reflected in clusters of normal network activity. 

As discussed in Section \ref{sec:detection}, we also propose an automatic choice of $k$ via \emph{QuickShift++}, 
in which case the two fast methods are denoted OC-Nystr\"om-QS, and OC-KJL-QS; \emph{these versions of our fast methods therefore only leave the choice of bandwidth $h$, and will be our main focus onwards}. 
\end{itemize}

Next, we discuss how the above parameters are picked in each of the use-cases or scenarios discussed above. 

\emph{- {\bf Minimal Tuning}: Validation of Hyperparameters.} To simulate the first training scenario where some small amount of labeled novelty data is available, we subsample a small amount of the novelty data (that is 75), which paired with equal amount of normal data is used to form a \emph{validation set} of size 150 to be used in hyperparamter choice; altogether, validation data sizes are kept very small relative to normal training size $n = 5000$. 

We then proceed to choosing $h$ or $k$ (when Quickshift++ is not used) to minimize AUC over the validation data, so that these choices are independent of the random test set on which final results are reported.  

\emph{- {\bf No Tuning}: Default Choice of Hyperparameters.} In this case, we choose the bandwidth $h$ by a common rule-of-thumb as the 0.25 quantile of increasing interpoint distances on the training data. The choice of number of components $k$ is then always made by Quickshift++. 

\paragraph{All Other Algorithmic Parameter Choices are Fixed}
We now describe all other choices inherent in our procedures, OC-Nystr\"om and OC-KJL, 
and their variants OC-Nystr\"om-QS, and OC-KJL-QS.

\emph{- Projection Parameters.} As discussed in Section \ref{sec:detection}, all projection parameters, namely subsamples size $m$, and projection dimension $d$, are fixed to $m = 100$ and $d = 5$, choices which work remarkably well in preserving detection performance across all datasets and types of novelty, despite the considerable amount of \emph{information compression} they entail.

\emph{- Quickshift++ Parameters.} 
We use the implementation of \cite{jiang2018quickshift++, quickshfit2018implementataion}, which requires internal parameters $\beta$ set to 0.9 (this performs density \emph{smoothing}) and number of neighbors set to $n^{2/3}$ (to build a \emph{dense} neighborhood graph whose connectivity encodes high-density regions), two choices which work well across device datasets and types of novelty. 

Here, due to variability in the data, Quickshift++ can often return too many \emph{outlier} clusters (despite the conservative setting of its internal parameters). To remove those, we only retain \emph{large} clusters, namely the smallest number of clusters that account for at least 95\% of the data, if this number is less than 20, otherwise we retain the 20 largest clusters discovered by Quickshift++.

\emph{Gaussian Mixture Models Parameters.} We have the choice of using either \emph{full} Gaussian covariances in fitting a GMM model to the projected data after KJL or Nystr\"om, or of using only diagonal covariances for faster fitting -- especially when operating in high dimensional settings -- but at the usual cost of some loss in accuracy. Since GMMs are fit after projection to low dimension $d=5$, it turns out that full Gaussian covariances are in fact efficient to fit in our case, so we only report results for full covariances. 

When using Quickshift++, we initialize GMM with the clusters returned, i.e., local means and covariances of these clusters, and train till convergence.

\section{Results to be Reported} \label{sec:results}

\subsection{Performance Metrics}\label{par:metrics}

\begin{itemize} [leftmargin=*]
\item {\bf Detection performance.} In novelty detection, there is a well known tension between \emph{false detection a.k.a. false positive rates} (FDR, i.e., the proportion of normal data wrongly flagged as novel) -- and \emph{true detection a.k.a. true positive rates} (TDR, i.e., the percentage of abnormal data rightly flagged as novel). 

This is because novelty detection consists of flagging a new observation $x$ as novel if it
scores below a given threshold $t$ (i.e., if $S(x) < t$ for some scoring function $S$ that is typically high for normal data, i.e., as $S(x)$ typically encodes some notion of \emph{similarity} to previous normal observations). Now, the higher the threshold $t$, the more likely it is that most novel
$x$'s -- most of which score much lower than $t$ -- are correctly detected (a true detection), but unfortunately, the more likely it is also that
normal datapoints are flagged as novel (a false detection), as many normal points will score high but lower than $t$. Thus, the best
performing detection procedures are those that alleviate this tension, i.e., achieving
high detection rates while minimizing false detection.

Such tradeoffs are well captured by a Receiver Operating Characteristic (ROC)
curve, which plots the detection rate TDR against the false alarm rate FDR as the detection threshold $t$ is varied from small to large; thus, the area under the ROC curve -- termed {\bf Area-Under-the-Curve (AUC)} -- when it is large, i.e., close to 1, indicates that good tradeoffs are achieved by the given detection approach. In contrast, AUC below 0.5 signals poor tradeoffs. 
AUC is therefore commonly adopted as a sensible measure of detection performance, as it captures the full performance tradeoff under the complete range of detection choices.

In practice, a single threshold is chosen, driven by application specific constraints, as one might prefer high TDR over low FDR, or the other way around (think for instance of an infected medical device, e.g., a pacemaker, where high TDR would be preferred, vs an infected smart home appliance, e.g., a toaster, where low FDR might be preferred). Large AUC, thus indicates that the detector allows for good choices in any of these situations. 

For our proposed fast detectors, we will be interested in the fraction of AUC retained over OCSVM, i.e., the AUC of our detector divided by that of OCSVM. 


\item {\bf Training and Detection (or Testing) Time.} We will measure time as the \emph{wall-clock time} taken by any of the methods for training (not-including data preprocessing into feature vectors, but inclusive of all actual training, i.e., modeling fitting), and \emph{testing}, i.e., actual detection computations, on given machine environments (see Section \ref{par:machine} below), after a model is obtained. 

For our proposed detectors, we will be reporting the \emph{time speedup}, i.e., the ratio of wall-clock time for our detector over that of OCSVM, separtely for training and testing. We empahsize that we are primarily interested in test time speedup, but need to ensure that our detectors remain practical to train. 

\item {\bf Detection (or Testing) Space.} We report the space taken by the model returned by the detection procedure in kiloBytes. Namely, we report the minimal amount of information on the learned model to be saved towards future detection. That is, (1) support vectors and coefficients for OCSVM, and (2) projection parameters and GMM components for OC-Nystr\"om and OC-KJL (with or without Quickshift++), all as described in Section \ref{sec:OCSVM} and \ref{sec:detection}. 


{While testing space is dependent on the programming language, in our case Python 3.7.3, space is machine-independent, as Python allows portability across 64 or 32 bits machine architectures via its \emph{pickling} process \cite{pickle}. All our models are first trained on a 64 bit server (Section \ref{par:machine}).}  

For our proposed detectors, we will be reporting the \emph{reduction} in space over OCSVM, i.e., the ratio of testing space of OCSVM over that of our cheaper detection procedures.  

\end{itemize}

\subsection{Computing Platforms}\label{par:machine}

 We consider two main computing platforms corresponding to important use cases: (1) a large and fast server -- for the use case where all training and detection happen offline  outside perhaps multiple IoT networks being monitored -- and (2) resource constrained nanodevices such as a Raspberry Pi or a router, corresponding to the use case where detection is to be realtime on the same IoT network being monitored. Details are given in Table \ref{tab:machines}.

\begin{table}[!t]
\centering
\caption{We train on server and test on all 3 machines}
\label{tab:machines}

 \begin{tabular}{| m{7em} | m{0.75\textwidth}|} 
 \hline

 \vspace{0.1cm}
 Machine & Description   \\ [0.5ex] 
 \hline\hline
 Large Server & 64-bit, running Debian GNU/Linux 9 (stretch) with Intel(R) Xeon(R) processor (32 CPU Cores, 1200-3400 MHz each), 100GB memory, and 2TB disk. \\
 
 \hline \hline 
 Raspberry Pi & 32-bit, running Raspbian GNU/Linux 10 (buster) with Cortex-A72 processor (4 CPU cores, 600-1500 MHz each), 8GB memory, and 27GB disk \\
 
 \hline 
 Nvidia Nano &  64-bit, running Ubuntu 18.04.5 LTS (Bionic Beaver) with Cortex-A57 processor (4 CPU cores, 102-1479 MHz each), 4GB memory, and 30GB disk\\
  [1ex] 
 \hline
 \end{tabular}
\end{table}





\subsection{Results Averaging}
To reduce uncertainty in reported results, we introduce repetitions in various stages of our experiments and report averages and standard deviations on performance metrics. 

For each dataset, first all flows (normal and abnormal) are preprocessed into the IAT+SIZE features described above in the experimental setup Section \ref{sec:experimentalsetup}. This results in a large pool of \emph{normal} and {novelty} data (see Table \ref{tab:dataset_details} of the Appendix), from which we we draw random subsamples. 

Experiments on each datset follow the steps outlined below. 
\begin{tcolorbox}[colframe=gray,boxsep=3pt,left=3pt,right=3pt,top=3pt,bottom=3pt]
\small 
\begin{enumerate} [leftmargin=*]
\item [(i)] Draw a subsample of size 300 from the normal pool, and a subsample of size 300 from the novelty pool to form a test dataset of size 600. 
\item [(ii)] Repeat 5 times for accurate AUC: 
\begin{itemize}[leftmargin=*]
\item Draw a subsample of size $n=5000$ from the normal pool, to form the training data. 
\item {\it If tuning:} draw a validation sample of size 150. 
\item Choose parameters $h, k$ as described in Section \ref{sec:experimentalsetup}.
\item Train with the choice of $h, k$ and save model on disk. 
\item Load model and test on Test data: repeat this 20 times for accurate timing on target machine (retain aggregate time). 
\end{itemize}
\end{enumerate} 

\end{tcolorbox}

Now for the baseline OCSVM, we simply report average and std of performance metrics over the 5 repetitions. When reporting \emph{speedups} for OC-Nystr\"om and OC-KJL over OCSVM, we use the corresponding average performance of OCSVM, say $\mu$. In other words, if we observe AUCs $a_1, \ldots, a_5$ for OC-KJL, we report the mean of $a_1/\mu, \ldots a_5/\mu$ $\pm$ the std of these ratios. We proceed similarly for time ratios. 

\section{Results Under Minimal Tuning}\label{sec:minimaltuning}
In this section, we consider the first situation where some validation is available to tune model hyperparmeters as described in Section \ref{sec:experimentalsetup}. We will see that as desired, OC-Nystr\"om and OC-KJL, with or without Quickshift++, indeed maintain the detection performance and training times of OCSVM, while significantly reducing testing time and space.

As previously discussed in Section \ref{sec:experimentalsetup}, all testing procedures are implemented in Python \texttt{numpy}, with parallelism turned on to take advantage of multicore systems. 

\begin{table*}[!htbp]
\VLS
\caption{OCSVM baseline performance. Time is in milliseconds per 100 datapoints and space is in kiloBytes. }
\label{tab:ocsvm_baseline}
\begin{tabular}{|F{1.2cm}|F{1cm}|F{1.9cm}|F{1.9cm}|F{1.9cm}|F{1.9cm}|F{1.9cm}|F{1.9cm}|F{1.9cm}|} 
\toprule
\multicolumn{2}{|c|}{Dataset}    & UNB      & CTU  & MAWI         & MACCDC       & SFRIG  & AECHO   &DWSHR  \\
\midrule
\multicolumn{2}{|c|}{AUC} &0.62 $\pm$ 0.03 & 0.65 $\pm$ 0.01 & 0.99 $\pm$ 0.00 & 0.86 $\pm$ 0.01 & 0.93 $\pm$ 0.00 & 0.91 $\pm$ 0.00 & 0.71 $\pm$ 0.01 \\
\midrule
\multicolumn{2}{|c|}{\C{Server Train \\Time (ms)}} & 44.54 $\pm$ 1.01  & 37.47 $\pm$ 0.82  & 68.05 $\pm$ 2.67  & 38.06 $\pm$ 1.16  & 37.84 $\pm$ 1.82  & 47.01 $\pm$ 1.26  & 37.17 $\pm$ 0.87  \\
\midrule
\multirow{3}{*}{\C{\hspace{-0.15cm}Test\\ Time\\ (ms)}} &\C{RSPI} & 65.69 $\pm$ 2.02 & 76.55 $\pm$ 2.32 & 89.54 $\pm$ 1.32 & 81.10 $\pm$ 1.69 & 80.72 $\pm$ 2.05 & 83.76 $\pm$ 1.09 & 81.95 $\pm$ 1.00 \\
\cmidrule{2-9}
&\C{NANO} & 39.24 $\pm$ 0.33 & 43.05 $\pm$ 0.14 & 50.58 $\pm$ 2.97 & 43.65 $\pm$ 0.47 & 41.01 $\pm$ 4.12 & 45.89 $\pm$ 0.13 & 41.55 $\pm$ 0.11 \\
\cmidrule{2-9}
&Server & 11.64 $\pm$ 0.16 & 12.71 $\pm$ 0.09 & 12.76 $\pm$ 0.73 & 13.08 $\pm$ 0.28 & 12.65 $\pm$ 1.57 & 13.05 $\pm$ 0.11 & 12.63 $\pm$ 0.08 \\
\midrule
\multicolumn{2}{|c|}{\C{Space (kB)}} & \C{974.87 $\pm$\\ 1.50} & \C{481.24 $\pm$\\ 0.29} & \C{2444.29 $\pm$\\ 5.50} & \C{521.92 $\pm$\\ 0.70} & \C{481.43 $\pm$\\ 0.31} & \C{1044.58 $\pm$\\ 0.68} & \C{441.89 $\pm$\\ 0.24} \\

\bottomrule
\end{tabular}
\end{table*}
\begin{table*}[!ht]
\VLS
\caption{Retained AUC (method/OCSVM) and server train time speedup
(OCSVM time/method time).}
\label{tab:iat_size-best-retained}
\begin{tabular}{|F{2.5cm}|F{1.9cm}|F{1.9cm}|F{1.9cm}|F{1.9cm}|F{1.9cm}|F{1.9cm}|F{1.9cm}|} 
\toprule
\diagbox[width=2.4cm,height=.6cm]{\hspace{-.2cm}Method}{\raisebox{1.27cm}{\hspace{3.cm}{Dataset}}}    & UNB      & CTU  & MAWI         & MACCDC       & SFRIG  & AECHO   &DWSHR  \\
\midrule
\C{\hspace{-1.3cm}OC-KJL:\\\hspace{.0cm}\uline{AUC Retained}} & 1.42 $\pm$ 0.03 & 1.15 $\pm$ 0.07 & 0.99 $\pm$ 0.02 & 1.08 $\pm$ 0.03  & 1.00 $\pm$ 0.01 & 1.06 $\pm$ 0.01 & 1.01 $\pm$ 0.02 \\
\cmidrule{2-8}
\RA{Train Speedup} & 1.98 $\pm$ 0.04 & 2.24 $\pm$ 0.05 & 3.82 $\pm$ 0.15 & 2.02 $\pm$ 0.06 & 2.19 $\pm$ 0.11 & 2.35 $\pm$ 0.06 & 1.96 $\pm$ 0.05 \\
\midrule
\C{\hspace{-.7cm}OC-KJL-QS:\\\hspace{-0.1cm}\uline{AUC Retained}} &1.41 $\pm$ 0.04 & 1.06 $\pm$ 0.04 & 0.91 $\pm$ 0.05 & 1.01 $\pm$ 0.02  & 1.00 $\pm$ 0.01 & 1.04 $\pm$ 0.02 & 0.98 $\pm$ 0.01 \\
\cmidrule{2-8}
\C{Train Speedup} & 1.23 $\pm$ 0.03 & 1.03 $\pm$ 0.02 & 1.88 $\pm$ 0.07 & 1.03 $\pm$ 0.03 & 1.03 $\pm$ 0.05 & 1.27 $\pm$ 0.03 & 1.00 $\pm$ 0.02 \\
\midrule
\hline 
\midrule
\C{\hspace{-.7cm}OC-Nystr\"om:\\\hspace{-0.1cm}\uline{AUC Retained}} &1.56 $\pm$ 0.01 & 1.35 $\pm$ 0.05 & 0.98 $\pm$ 0.02 & 1.08 $\pm$ 0.02  & 0.98 $\pm$ 0.02 & 1.06 $\pm$ 0.01 & 1.04 $\pm$ 0.01 \\
\cmidrule{2-8}
\C{Train Speedup} &2.56 $\pm$ 0.06 & 2.20 $\pm$ 0.05 & 3.74 $\pm$ 0.15 & 2.05 $\pm$ 0.06 & 2.30 $\pm$ 0.11 & 2.50 $\pm$ 0.07 & 1.97 $\pm$ 0.05 \\
\midrule
\C{\hspace{-0.1cm}OC-Nystr\"om-QS:\\\hspace{-0.1cm}\uline{AUC Retained}} & 1.55 $\pm$ 0.01 & 1.20 $\pm$ 0.06 & 0.96 $\pm$ 0.02 & 1.04 $\pm$ 0.04  & 1.00 $\pm$ 0.01 & 1.05 $\pm$ 0.01 & 0.99 $\pm$ 0.01 \\ 
\cmidrule{2-8}
\C{Train Speedup} &1.04 $\pm$ 0.02 & 1.02 $\pm$ 0.02 & 1.88 $\pm$ 0.07 & 1.03 $\pm$ 0.03 & 1.06 $\pm$ 0.05 & 1.23 $\pm$ 0.03 & 0.95 $\pm$ 0.02 \\
\bottomrule
\end{tabular}
\end{table*}

\subsection{OCSVM Baseline Performance}
Table \ref{tab:ocsvm_baseline} provides OCSVM baseline performance results. All training is performed on the server, while we test on all 3 machines. The table reports (1) AUC, same for all machines, since the same models and test data are used for fair comparison, (2) training time on the server, and (3) test time for all 3 machines and (4) test space, again same for all machines.

\subsection{Retained AUC and Training Efficiency}
As stated earlier, we now verify that our proposed methods manage to retain the accuracy of the baseline OCSVM, and also do not sacrifice training efficiency. These are the results of Table \ref{tab:iat_size-best-retained}.

\paragraph{AUC retained}  We see that our detection methods OC-Nystr\"om and OC-KJL, with or without Quickshift++, largely retain the detection performance of OCSVM, all within a ratio of $1$ or more, except in the case of MAWI with OC-KJL-QS where 91$\pm$ 5\% of OCSVM's AUC is retained  -- which is still a high AUC considering OCSVM is nearly perfect on MAWI.  
. Moreover, for some datasets such as UNB and CTU, all procedures manage to actually outperform OCSVM, even quite significantly in the case of UNB. It is likely that such higher performance is due to the additional regularization inherent in the dimension reduction performed by our methods. 

We also remark that the versions with Quickshift++, namely OC-Nystr\"om-QS and OC-KJL-QS, tend to achieve slightly smaller AUC compared to the non-Quickshift++ counterparts where the number of components $k$ is tuned by validation. However, as already stated, they also manage to maintain or outperform the basline AUC of OCSVM. 

\paragraph{Training time} Although our original goal was just to maintain the training  efficiency as OCSVM, especially considering the various additional steps inherent in our methods, our methods without Quickshift++ in fact achieve speedup --  factors of 2-3 in some cases -- over OCSVM training time which involves more expensive model fitting steps.  In the case of OC-Nystr\"om-QS and OC-KJL-QS, training time is slower due to the Quickshift automatic search for the right number of clusters in the projected data. 

Henceforth, in the rest of the main paper body, we will focus attention to OC-Nystr\"om-QS and OC-KJL-QS, as their performance is similar to their non Quickshift++ counterparts, while at the same time they are more readily applicable in all scenarios, including when no validation data is available for tuning (results of Section \ref{sec:notuningresults}).

\subsection{Significant Savings in Detection Time and Space}
This is perhaps the most important section of this work, where we report significant savings on detection time and space, for all proposed variants of our approach, which is the main motivation of this work. 

Table \ref{tab:iat_size-best-oc_kjl_qs} and Table \ref{tab:iat_size-best-oc_nystrom_qs} present results for OC-Nystr\"om-QS, and OC-KJL-QS, while similar time and space savings under OC-Nystr\"om and OC-KJL are presented in Appendix \ref{sec:moreOCresults}, Tables \ref{tab:iat_size-best-oc_kjl} and \ref{tab:iat_size-best-oc_nystrom}. 

\paragraph{Testing time speedup} We observe that our approaches are at least
9.5 times faster than OCSVM on every machine we considered, Nvidia Nano,
Raspberry Pi, and the server. Speedups on Raspberry Pi and the server are most
considerable, up to 20+ times faster than OCSVM on many datasets.  The smaller
amount of speedup that we observe on the Nano can be attributed to the
relatively smaller amount of memory that this device has compared to the
Raspberry Pi, which likely forces more memory swap operations as all test data
is loaded in at once. We also note that unlike the Nano and Raspberry Pi, the
server may have had more competing processes, yet even on the server, the
trend of large speedups is observed across datasets. 

Finally, here we see a small distinction between OC-KJL-QS and
OC-Nystr\"om-QS, whereby the former tends to achieve higher speedups on all
machines for most datasets. As such both approaches seem to offer a tradeoff
where, as per Table \ref{tab:iat_size-best-retained}, the Nystr\"om based
approaches tend to achieve slightly higher AUC on most datasets. 

\paragraph{Space reduction} In all cases we observe significant space
reductions -- this is machine independent -- as our models can be stored
upwards of 17 times less space, and up to 23 times less that the baseline
OCSVM model.  This smaller memory footprint imples the possibility for a much
wider deployment than a conventional OCSVM, especially on memory restricted
devices such as the embedded devices on which we conducted our evaluation.
Although we focused much of our evaluation on memory constrained devices,
which is a common deployment scenario for IoT, the space efficiency of these
models is important even in server settings where a server might host large
numbers of detection tools each dedicated to monitoring a given machine on
client networks.

\section{Results under No Tuning}\label{sec:notuningresults}
We now consider the scenario where no validation data is available to tune any of the procedures, i.e., in choosing the bandwidth parameter $h$. While in general it is preferable to perform some minimal tuning before deployment, in practice it may be difficult to obtain labeled data for the types of novel activities of interest that commonly arise in actual deployment environments.

In the practice of novelty detection with OCSVM, when no labelled data is available, various rule-of-thumbs are used, a popular one being to pick $h$ as a quantile of interpoint distances. For uniformity, as explained in Section \ref{sec:experimentalsetup}, here we pick $h$ for all methods, as the 25th percentile of increasing interpoint distances in the training data. 

Naturally, detection performance suffers w.r.t. that of a tuned procedure, for any of the methods. Furthermore, since the choice of bandwidth affect the learned model, it is to be expected that time and space comparisons would also differ from that under minimal tuning as in the previous Section \ref{sec:minimaltuning}.

\begin{table*}[!ht]
\VLS
\caption{OC-KJL-QS: Test time speedup
(OCSVM over method) and space reduction (OCSVM over method).}
\label{tab:iat_size-best-oc_kjl_qs}
\begin{tabular}{|F{1.2cm}|m{1cm}|F{1.9cm}|F{1.9cm}|F{1.9cm}|F{1.9cm}|F{1.9cm}|F{1.9cm}|F{1.9cm}|} 
\toprule
\multicolumn{2}{|c|}{Dataset}    & UNB      & CTU  & MAWI         & MACCDC       & SFRIG  & AECHO   &DWSHR  \\
\midrule
\multirow{3}{*}{\C{\hspace{-0.15cm}Test\\ Time\\Speedup}} &\C{RSPI} & 22.54 $\pm$ 0.69 & 19.68 $\pm$ 0.60 & 21.52 $\pm$ 0.32 & 20.97 $\pm$ 0.44 & 19.54 $\pm$ 0.50 & 20.48 $\pm$ 0.27 & 19.35 $\pm$ 0.24 \\
\cmidrule{2-9}
&\C{NANO} & 11.82 $\pm$ 0.10 & 12.08 $\pm$ 0.04 & 14.79 $\pm$ 0.87 & 13.25 $\pm$ 0.14 & 10.97 $\pm$ 1.10 & 12.97 $\pm$ 0.04 & 11.44 $\pm$ 0.03 \\
\cmidrule{2-9}
&\C{Server} & 17.15 $\pm$ 0.24 & 16.66 $\pm$ 0.12 & 19.52 $\pm$ 1.12 & 21.63 $\pm$ 0.46 & 19.76 $\pm$ 2.45 & 21.05 $\pm$ 0.18 & 18.39 $\pm$ 0.11 \\
\midrule
\multicolumn{2}{|c|}{\C{Space Reduction}} &22.02 $\pm$ 0.03 & 18.46 $\pm$ 0.01 & 23.80 $\pm$ 0.05 & 19.38 $\pm$ 0.03 & 17.88 $\pm$ 0.01 & 21.62 $\pm$ 0.01 & 17.59 $\pm$ 0.01 \\
\bottomrule
\end{tabular}
\end{table*}

\begin{table*}[!ht]
\VLS
\caption{OC-Nystr\"om-QS: Test time speedup
(OCSVM over method) and space reduction (OCSVM over method).}
\label{tab:iat_size-best-oc_nystrom_qs}
\begin{tabular}{|F{1.2cm}|m{1cm}|F{1.9cm}|F{1.9cm}|F{1.9cm}|F{1.9cm}|F{1.9cm}|F{1.9cm}|F{1.9cm}|} 
\toprule
\multicolumn{2}{|c|}{Dataset}   & UNB      & CTU  & MAWI         & MACCDC       & SFRIG  & AECHO   &DWSHR  \\
\midrule
\multirow{3}{*}{\C{\hspace{-0.15cm}Test\\ Time\\Speedup}} &\C{RSPI} &17.99 $\pm$ 0.55 & 17.74 $\pm$ 0.54 & 21.50 $\pm$ 0.32 & 22.10 $\pm$ 0.46 & 18.78 $\pm$ 0.48 & 19.49 $\pm$ 0.25 & 19.08 $\pm$ 0.23 \\
\cmidrule{2-9}
&\C{NANO} & 9.52 $\pm$ 0.08  & 11.29 $\pm$ 0.04 & 14.62 $\pm$ 0.86 & 13.67 $\pm$ 0.15 & 11.14 $\pm$ 1.12 & 12.49 $\pm$ 0.03 & 11.28 $\pm$ 0.03 \\
\cmidrule{2-9}
&\C{Server} &13.85 $\pm$ 0.20 & 15.14 $\pm$ 0.11 & 22.05 $\pm$ 1.26 & 21.21 $\pm$ 0.45 & 15.53 $\pm$ 1.93 & 16.47 $\pm$ 0.14 & 16.02 $\pm$ 0.10 \\
\midrule
\multicolumn{2}{|c|}{\C{Space Reduction}} &20.62 $\pm$ 0.03 & 17.80 $\pm$ 0.01 & 23.82 $\pm$ 0.05 & 19.62 $\pm$ 0.03 & 17.98 $\pm$ 0.01 & 21.46 $\pm$ 0.01 & 17.41 $\pm$ 0.01 \\
\bottomrule
\end{tabular}
\end{table*}

\subsection{Baseline OCSVM Performance}
Table \ref{tab:iat_size-default-ocsvm_baseline} shows the performance of the baseline OCSVM. We observe a small decrease in AUC for most datasets, most considerably for UNB and CTU which already were hard datasets even under tuning (Table \ref{tab:ocsvm_baseline}). Interestingly, MAWI, SFRIG. AECHO and MACCDC still admit high AUCs even without tuning, attesting to the general appeal of OCSVM as an adaptable and robust novelty detection approach.  

\subsection{Retained AUC and Training Efficiency}
Table \ref{tab:iat_size-default-retained} compares AUC and training times of  
OC-Nystr\"om-QS and OC-KJL-QS to that of the baseline OCSVM, using the exact same default choice of bandwidth $h$ as OCSVM. 

\paragraph{AUC retained} OC-Nystr\"om-QS and OC-KJL-QS manage to retain the AUC of OCSVM on most datasets. However, on MAWI, neither OC-Nystr\"om-QS nor OC-KJL-QS does well, arriving at just a fraction of the baseline AUC. SFRIG also appears to cause problems for OC-Nystr\"om-QS under the default $h$ setting. Interestingly, both approaches again outperform the baseline on UNB and CTU, and less significantly so on DWSHR.  

\paragraph{Training time}

As before, training time remains competitive with that of OCSVM, with some significant reduction in time for instance in the case of UNB and AECHO.

\begin{table*}[!htbp]
\VLS
\caption{OCSVM baseline performance, {\bf no tuning}. Time is in ms per 100 datapoints and space is in kB. }
\label{tab:iat_size-default-ocsvm_baseline}
\begin{tabular}{|F{1.3cm}|F{1cm}|F{1.9cm}|F{1.9cm}|F{1.9cm}|F{1.9cm}|F{1.9cm}|F{1.9cm}|F{1.9cm}|} 
\toprule

\multicolumn{2}{|c|}{Dataset}    & UNB      & CTU  & MAWI         & MACCDC       & SFRIG  & AECHO   &DWSHR  \\
\midrule
\multicolumn{2}{|c|}{AUC} &0.59 $\pm$ 0.00 & 0.59 $\pm$ 0.02 & 0.99 $\pm$ 0.00 & 0.81 $\pm$ 0.03 & 0.93 $\pm$ 0.00 & 0.85 $\pm$ 0.00 & 0.68 $\pm$ 0.01 \\
\midrule
\multicolumn{2}{|c|}{\C{Server Train \\Time (ms)}} & 45.82 $\pm$ 0.68  & 37.70 $\pm$ 0.86  & 65.41 $\pm$ 0.57  & 38.81 $\pm$ 0.91  & 39.12 $\pm$ 0.26  & 49.61 $\pm$ 0.72  & 37.68 $\pm$ 0.43  \\
\midrule
\multirow{3}{*}{\C{\hspace{-0.15cm}Test\\ Time\\ (ms)}} &\C{RSPI} & 74.07 $\pm$ 1.87 & 80.56 $\pm$ 1.28 & 89.40 $\pm$ 1.50 & 80.90 $\pm$ 1.47 & 82.88 $\pm$ 1.00 & 83.28 $\pm$ 1.38 & 81.16 $\pm$ 1.40 \\
\cmidrule{2-9}
&\C{NANO} & 44.66 $\pm$ 0.13 & 43.04 $\pm$ 0.13 & 44.37 $\pm$ 0.10 & 43.83 $\pm$ 0.11 & 44.52 $\pm$ 0.42 & 45.74 $\pm$ 0.12 & 43.58 $\pm$ 0.17 \\
\cmidrule{2-9}
&\C{Server} & 13.01 $\pm$ 0.09 & 12.85 $\pm$ 0.12 & 11.11 $\pm$ 0.06 & 12.89 $\pm$ 0.09 & 14.15 $\pm$ 0.38 & 12.78 $\pm$ 0.27 & 13.00 $\pm$ 0.06 \\
\midrule
\multicolumn{2}{|c|}{\C{Space (kB)}} & \C{962.51 $\pm$\\ 0.42} & \C{481.39 $\pm$\\ 0.08} & \C{2447.42 $\pm$\\ 1.07} & \C{521.54 $\pm$\\ 0.28} & \C{481.20 $\pm$\\ 0.14} & \C{1042.33 $\pm$\\ 0.31} & \C{441.04 $\pm$\\ 0.18} \\

\bottomrule
\end{tabular}
\end{table*}
\FloatBarrier

\subsection{Significant Savings in Detection Time and Space}
Tables \ref{tab:iat_size-default-oc_kjl_qs} and \ref{tab:iat_size-default_oc_nystrom_qs} presents results on detection time and space savings for both OC-Nystr\"om-QS and OC-KJL-QS, again with the same default choice of bandwidth $h$ as OCSVM. The trends on savings are similar, but in fact even better than those under minimal tuning of these 3 methods. 

\paragraph{Testing time speedup} We observe speedups of at least 10.6 times over the baseline OCSVM detection times across all machines and datasets. Again, the most speedups are observed on Raspberry Pi and the server, while the smaller memory Nano tends to achieve smaller but still significant speedups. 

Finally, we again observe the trend where OC-KJL-QS manages faster times than OC-Nystr\"om-QS in most cases, especially on the Raspberry and server machines. 

\paragraph{Space reduction} As before, space reductions are significant w.r.t. to the baseline OCSVM, from 16 to 20+ times less space than required by the baseline. 

\section{Conclusion}
\subsection{Summary}
The very nature of IoT devices, namely the fact that they tend to have few modes of operations, makes it possible to succinctly model their normal behavior, and in particular \emph{model} their network flows into relatively few \emph{clusters} of activity, under appropriate representations of the data. Here, starting with the very predictive data representation achieved by OCSVM, we can reduce the vanilla OCSVM model to more efficient representations by projection and clustering to achieve remarkable savings in novelty detection tasks, both in terms of time and space, and this without sacrificing detection accuracy. The resulting approaches, OC-Nystr\"om and OC-KJL are therefore more widely applicable under practical use cases of novelty detection in IoT, and in particular deployable not only on powerful servers -- as is usually the case with machine learning procedures -- but also on nano-computing devices with more limited memory and computing resources. 

Our two main approaches offer some visible tradeoffs as argued above: when minimally tuned with a few labeled data, OC-Nystr\"om tends to achieve higher detection performance than OC-KJL, however at the cost of some decrease in time efficiency. This is the case under both versions of these approaches, i.e., with or without automatic cluster-number detection with Quickshift++. 

Under default settings, where we always use Quickshift++, OC-KJL is to be preferred as it achieves noticeably faster detection time, but maintains similar and sometimes better accuracy than  OC-Nystr\"om.



\begin{table*}[!ht]
\VLS
\caption{{\bf No tuning.} Retained AUC (method over OCSVM) and train time speedup
(OCSVM over method).}
\label{tab:iat_size-default-retained}
\begin{tabular}{|F{2.6cm}|F{1.9cm}|F{1.9cm}|F{1.9cm}|F{1.9cm}|F{1.9cm}|F{1.9cm}|F{1.9cm}|} 
\toprule
\diagbox[width=2.6cm,height=.6cm]{\raisebox{-0.02cm}{\vspace{0.0cm}Method}}{\raisebox{0.5cm}{\vspace{-0.3cm}\hspace{-1.0cm}Dataset}}    & UNB      & CTU  & MAWI         & MACCDC       & SFRIG  & AECHO   &DWSHR  \\
\midrule
\C{\hspace{-0.8cm}OC-KJL-QS:\\\hspace{0.3cm}\uline{AUC Retained}} &1.48 $\pm$ 0.06 & 1.14 $\pm$ 0.02 & 0.13 $\pm$ 0.00 & 1.01 $\pm$ 0.07  & 0.83 $\pm$ 0.06 & 0.96 $\pm$ 0.09 & 1.04 $\pm$ 0.01 \\
\cmidrule{2-8}
\C{\hspace{0.3cm}Train Speedup} & 1.28 $\pm$ 0.02  & 1.06 $\pm$ 0.02  & 1.79 $\pm$ 0.02  & 1.11 $\pm$ 0.03  & 1.05 $\pm$ 0.01  & 1.32 $\pm$ 0.02  & 0.99 $\pm$ 0.01  \\
\midrule
\hline 
\C{\hspace{-0.2cm}OC-Nystr\"om-QS:\\\hspace{0.3cm}\uline{AUC Retained}} & 1.58 $\pm$ 0.06 & 1.23 $\pm$ 0.07 & 0.16 $\pm$ 0.00 & 1.05 $\pm$ 0.06  & 0.42 $\pm$ 0.10 & 1.09 $\pm$ 0.06 & 0.96 $\pm$ 0.03 \\ 
\cmidrule{2-8}
\C{\hspace{0.3cm}Train Speedu}p &1.04 $\pm$ 0.02  & 1.01 $\pm$ 0.02  & 1.67 $\pm$ 0.01  & 1.08 $\pm$ 0.03  & 1.03 $\pm$ 0.01  & 1.28 $\pm$ 0.02  & 0.96 $\pm$ 0.01 \\
\bottomrule
\end{tabular}
\end{table*}
\FloatBarrier

\begin{table*}[!htp]
\VLS
\caption{OC-KJL-QS, {\bf no tuning}. Test time speedup
(OCSVM over method) and space reduction (OCSVM over method).}
\label{tab:iat_size-default-oc_kjl_qs}
\begin{tabular}{|F{1.2cm}|m{1cm}|F{1.9cm}|F{1.9cm}|F{1.9cm}|F{1.9cm}|F{1.9cm}|F{1.9cm}|F{1.9cm}|} 
\toprule
\multicolumn{2}{|c|}{Dataset}    & UNB      & CTU  & MAWI         & MACCDC       & SFRIG  & AECHO   &DWSHR  \\
\midrule
\multirow{3}{*}{\C{\hspace{-0.15cm}Test\\ Time\\Speedup}} &  \C{RSPI} & 24.64 $\pm$ 0.62& 21.36 $\pm$ 0.34& 16.50 $\pm$ 0.28& 22.83 $\pm$ 0.42& 20.32 $\pm$ 0.25& 19.39 $\pm$ 0.32& 19.00 $\pm$ 0.33\\
\cmidrule{2-9}
& \C{NANO} & 14.07 $\pm$ 0.04 & 12.52 $\pm$ 0.04 & 10.60 $\pm$ 0.02 & 13.97 $\pm$ 0.04 & 12.82 $\pm$ 0.12 & 12.57 $\pm$ 0.03 & 11.91 $\pm$ 0.05\\
\cmidrule{2-9}
&\C{Server}&25.30 $\pm$ 0.17& 20.08 $\pm$ 0.19& 13.22 $\pm$ 0.07& 19.68 $\pm$ 0.14& 18.08 $\pm$ 0.49& 17.00 $\pm$ 0.36& 21.57 $\pm$ 0.11\\
\midrule
\multicolumn{2}{|c|}{\C{Space Reduction}}&22.03 $\pm$ 0.01 & 18.64 $\pm$ 0.00 & 22.99 $\pm$ 0.01 & 19.73 $\pm$ 0.01 & 18.56 $\pm$ 0.01 & 21.61 $\pm$ 0.01 & 17.73 $\pm$ 0.01\\

\bottomrule
\end{tabular}
\end{table*}
\FloatBarrier

\FloatBarrier
\begin{table*}[!htp]
\VLS
\caption{OC-Nystr\"om-QS, {\bf no tuning}. Test time speedup
(OCSVM over method) and space reduction (OCSVM over method).}
\label{tab:iat_size-default_oc_nystrom_qs}
\begin{tabular}{|F{1.2cm}|m{1cm}|F{1.9cm}|F{1.9cm}|F{1.9cm}|F{1.9cm}|F{1.9cm}|F{1.9cm}|F{1.9cm}|} 
\toprule
\multicolumn{2}{|c|}{Dataset}    & UNB      & CTU  & MAWI         & MACCDC       & SFRIG  & AECHO   &DWSHR  \\
\midrule
\multirow{3}{*}{\C{\hspace{-0.15cm}Test\\ Time\\Speedup}} &  \C{RSPI} &18.05 $\pm$ 0.46& 19.27 $\pm$ 0.31& 16.42 $\pm$ 0.28& 21.86 $\pm$ 0.40& 19.19 $\pm$ 0.23& 18.37 $\pm$ 0.31& 17.67 $\pm$ 0.30\\
\cmidrule{2-9}
& \C{NANO} &10.88 $\pm$ 0.03 & 11.82 $\pm$ 0.04 & 10.61 $\pm$ 0.02 & 13.94 $\pm$ 0.04 & 11.98 $\pm$ 0.11 & 12.06 $\pm$ 0.03 & 10.53 $\pm$ 0.04\\
\cmidrule{2-9}
&\C{Server} &15.84 $\pm$ 0.11& 17.29 $\pm$ 0.16& 13.96 $\pm$ 0.07& 14.65 $\pm$ 0.10& 20.76 $\pm$ 0.56& 16.84 $\pm$ 0.35& 19.13 $\pm$ 0.09\\
\midrule
\multicolumn{2}{|c|}{\C{Space Reduction}} &20.40 $\pm$ 0.01 & 17.94 $\pm$ 0.00 & 22.99 $\pm$ 0.01 & 19.65 $\pm$ 0.01 & 18.10 $\pm$ 0.01 & 21.26 $\pm$ 0.01 & 16.85 $\pm$ 0.01\\

\bottomrule
\end{tabular}
\end{table*}
\FloatBarrier

\balance
\subsection{Open Questions}

A main open question concerns practical deployment scenarios where model drift
occurs.  For example, a device's \emph{normal} behavior might change over
time, for a variety of reasons. In practical deployments, a device's normal
behavior might change over time, due to day of the week, weekday vs. weekend,
seasonally, and so forth. Furthermore, exogenous events such as software
upgrades may also cause changes to the underlying traffic chracteristics that
result in the incorrect detection of novel behavior.  Without taking such
drift into account, the current approach may incorrectly detect certain events
as novel. This outcome could potentially create false alarms until the system
is properly retrained with the new data containing previously unobserved
modalities. Such retraining can be expensive, and ideally, we would want a
system that can efficiently update itself in real-time, leveraging previous
models as additional data is acquired over time. Efficiently {\em updating}
the models we have developed for incremental updates as retraining becomes
necessary and new data becomes available.

From a practical standpoint, we might also consider whether the outputs these
models produce are actionable. While unsupervised learning techniques offer
the convenience of being able to train without labeled data, selecting detection thresholds
that are actionable for those running the detection algorithms may depend on
the specific circumstance. Finally, one might consider various aspects of data
representation, and how different representations might also improve the
efficiency of training. For example, one of the advantages of the models that
we have developed is that they operate directly on simple sets of features
derived from packet traces. Yet, other work has explored whether even simpler,
packet-based representations might be appropriate for certain
problems~\cite{holland2020nprint}; exploring how these representations (or
variations of them) perform with unsupervised models is another possible
future direction.



\clearpage

\bibliographystyle{abbrv}
\balance\bibliography{reference}

\appendix

\counterwithin{figure}{section} \counterwithin{table}{section}

\section{Dataset Pool Sizes} \label{dataset}

Here we describe the initial number of normal and abnormal flows in each of the datasets of Table \ref{tab:dataset_details}. As described in the main text these are subsampled from to form the training, validation and test data used in our experiments. All the initial sizes are given in Table \ref{tab:dataset_details}. 

\begin{table}[h]
\centering
\setlength{\extrarowheight}{0pt}
\addtolength{\extrarowheight}{\aboverulesep}
\addtolength{\extrarowheight}{\belowrulesep}
\setlength{\aboverulesep}{0pt}
\setlength{\belowrulesep}{0pt}
\caption{Dataset size.}
\label{tab:dataset_details}
\begin{tabular}{|F{1.2cm}|F{0.9cm}|F{0.9cm}|F{0.99cm}|F{1.5cm}|F{0.99cm}|F{1.2cm}|F{1.3cm}|}
\toprule
Dataset & UNB & CTU & MAWI &  MACCDC & SFRIG & AECHO  & DWSHR\\
\midrule
Normal  &   26942  & 21970  & 8320   &  24233  &  86088     &    27621  & 153089 \\
\midrule
Novelty  &   1284  & 6929  &  5558 & 5721  &    903  & 729    & 335  \\    
\bottomrule
\end{tabular}
\end{table}

\section{Minimal Tuning: OC-Nystr\"om and OC-KJL Savings}
\label{sec:moreOCresults}

In the main paper body, we left out some of the detection time and space savings results for the OC-Nystr\"om and OC-KJL variants (which don't use Quickshift++ for automatic cluster-number identification). These results are presented here in the appendix in Tables 
\ref{tab:iat_size-best-oc_kjl} and \ref{tab:iat_size-best-oc_nystrom}. 

We see that, just as the Quickshift++ variants, we observe significant speedups in detection time and space, with the most time speedups observed on Raspberry Pi and the server, which both have more memory space than the Nano. 

\FloatBarrier
\begin{table*}[!hbp]
\VLS
\caption{OC-KJL: Test time speedup
(OCSVM over method) and space reduction (OCSVM over method).}
\label{tab:iat_size-best-oc_kjl}
\begin{tabular}{|F{1.2cm}|m{1cm}|F{1.9cm}|F{1.9cm}|F{1.9cm}|F{1.9cm}|F{1.9cm}|F{1.9cm}|F{1.9cm}|} 
\toprule
\multicolumn{2}{|c|}{Dataset}    & UNB      & CTU  & MAWI         & MACCDC       & SFRIG  & AECHO   &DWSHR  \\
\midrule
\multirow{3}{*}{\C{\hspace{-0.15cm}Test\\ Time\\Speedup}} &  \C{RSPI} & 20.77 $\pm$ 0.64& 19.85 $\pm$ 0.60& 17.83 $\pm$ 0.26& 20.28 $\pm$ 0.42& 19.88 $\pm$ 0.50& 19.43 $\pm$ 0.25& 19.18 $\pm$ 0.23\\
\cmidrule{2-9}
& \C{NANO} & 10.87 $\pm$ 0.09 & 12.74 $\pm$ 0.04 & 12.02 $\pm$ 0.70 & 12.32 $\pm$ 0.13 & 11.25 $\pm$ 1.13 & 12.01 $\pm$ 0.03 & 11.38 $\pm$ 0.03\\
\cmidrule{2-9}
&\C{Server} & 17.52 $\pm$ 0.25& 19.36 $\pm$ 0.14& 16.38 $\pm$ 0.94& 17.32 $\pm$ 0.37& 11.94 $\pm$ 1.48& 16.65 $\pm$ 0.14& 12.67 $\pm$ 0.08\\
\midrule
\multicolumn{2}{|c|}{\C{Space Reduction}} & 21.53 $\pm$ 0.03 & 18.67 $\pm$ 0.01 & 23.22 $\pm$ 0.05 & 18.72 $\pm$ 0.03 & 17.98 $\pm$ 0.01 & 21.24 $\pm$ 0.01 & 17.49 $\pm$ 0.01\\
\bottomrule
\end{tabular}
\end{table*}
\FloatBarrier

\FloatBarrier
\begin{table*}[!hbp]
\VLS
\caption{OC-Nystr\"om: Test time speedup
(OCSVM over method) and space reduction (OCSVM over method).}
\label{tab:iat_size-best-oc_nystrom}
\begin{tabular}{|F{1.2cm}|m{1cm}|F{1.9cm}|F{1.9cm}|F{1.9cm}|F{1.9cm}|F{1.9cm}|F{1.9cm}|F{1.9cm}|} 
\toprule
\multicolumn{2}{|c|}{Dataset}   & UNB      & CTU  & MAWI         & MACCDC       & SFRIG  & AECHO   &DWSHR  \\
\midrule
\multirow{3}{*}{\C{\hspace{-0.15cm}Test\\ Time\\Speedup}} &  \C{RSPI} &22.42 $\pm$ 0.69& 20.05 $\pm$ 0.61& 18.95 $\pm$ 0.28& 20.68 $\pm$ 0.43& 23.11 $\pm$ 0.59& 19.91 $\pm$ 0.26& 18.55 $\pm$ 0.23\\
\cmidrule{2-9}
& \C{NANO} &11.83 $\pm$ 0.10 & 12.56 $\pm$ 0.04 & 12.95 $\pm$ 0.76 & 12.20 $\pm$ 0.13 & 13.00 $\pm$ 1.30 & 12.33 $\pm$ 0.03 & 10.69 $\pm$ 0.03\\
\cmidrule{2-9}
&\C{Server} &20.71 $\pm$ 0.29& 15.98 $\pm$ 0.12& 17.96 $\pm$ 1.03& 14.25 $\pm$ 0.30& 20.14 $\pm$ 2.50& 15.02 $\pm$ 0.13& 15.34 $\pm$ 0.09\\
\midrule
\multicolumn{2}{|c|}{\C{Space Reduction}} &22.06 $\pm$ 0.03 & 18.73 $\pm$ 0.01 & 23.37 $\pm$ 0.05 & 18.72 $\pm$ 0.03 & 19.42 $\pm$ 0.01 & 21.33 $\pm$ 0.01 & 17.01 $\pm$ 0.01\\
\bottomrule
\end{tabular}
\end{table*}
\FloatBarrier

\section{Alternative Features}

\subsection{SAMP-SIZE Features Description}
\label{samp_size}
SAMP-SIZE: a flow is partitioned into small time intervals of equal length, and the total packet size (i.e., byte count) in each interval is recorded; thus, a flow is represented as a time series of byte counts in small time intervals. Here, we obtain time intervals according to different quantiles (i.e., [0.1, 0.2, 0.3, 0.4, 0.5, 0.6, 0.7, 0.8, 0.9, 0.95]) of flow durations.
To ensure that each sample has the same dimension $D$, we select $D$ for all flows as the 90th percentile of all \emph{flow lengths} in the dataset (here, \emph{flow length} stands for the number of packets a flow -- as opposed to its duration in time).

Now for any given flow, if the number of fixed time intervals in the flow is less than $D$, we append 0's to arrive at a vector of dimension $D$. If instead the number of fixed time intervals is greater than $D$, we truncate the resulting vector representation down to dimension $D$. 

\subsection{STATS+HEADER Features Description}
\label{stats+header}
STATS+HEADER: a set of statistical quantities compiled from a flow. In particular, we choose 10 of the most common
such statistics in the literature (see e.g., \cite{moore2013discriminators}), namely, flow duration, number of packets sent per second,
number of bytes per second, and the following statistics on packet sizes (in bytes) in a flow: mean, standard
deviation, the first to third quantiles, the minimum, and maximum. Also, We incorporate packet header information (i.e., Time to Live (TTL) and TCP flags (FIN, SYN, RST, PSH, ACK, URG, ECE, and CWR) into the STATS to form the STATS+HEADER feature.

\subsection{Results under STATS+HEADER}
\label{appendix_stats+header}

\subsubsection{Results Under Minimal Tuning}

Table. \ref{tab:stats_header-best-ocsvm_baseline} shows the baseline results obtained by OCSVM under minimal turning.

Similar to the case of IAT+SIZE features, both OC-Nystr\"om and OC-KJL, with or without Quickshift++ retain the AUC and train time of the baseline OCSVM as shown in Table \ref{tab:stats_header-best-retained}.

We also see that these methods, under the alternative features attain significant detection time speedups over OCSVSM: this is shown in Tables \ref{tab:stats_header-best-oc_kjl_qs} and 
\ref{tab:stats_header-best-oc_nystrom_qs}. As in the main text, i.e., in the case of IAT+SIZE features, most significant speedups are obtained when running on Raspberry Pi and the server, which have more memory space than the Nano. 

\FloatBarrier
\begin{table*}[!hbp]
\VLS
\caption{OCSVM performance with STATS+HEADER. Time is in ms per 100 datapoints and space is in KB. }
\label{tab:stats_header-best-ocsvm_baseline}
\begin{tabular}{|F{1.2cm}|m{1cm}|F{1.9cm}|F{1.9cm}|F{1.9cm}|F{1.9cm}|F{1.9cm}|F{1.9cm}|F{1.9cm}|} 
\toprule
\multicolumn{2}{|c|}{Dataset}    & UNB      & CTU  & MAWI         & MACCDC       & SFRIG  & AECHO   &DWSHR  \\
\midrule
\multicolumn{2}{|c|}{AUC} &0.62 $\pm$ 0.00 & 0.60 $\pm$ 0.01 & 1.00 $\pm$ 0.00 & 0.74 $\pm$ 0.03 & 0.92 $\pm$ 0.01 & 0.97 $\pm$ 0.00 & 0.64 $\pm$ 0.00 \\
\midrule
\multicolumn{2}{|c|}{\C{Server Train \\Time (ms)}} & 39.88 $\pm$ 0.84  & 39.77 $\pm$ 0.69  & 54.48 $\pm$ 1.27  & 40.21 $\pm$ 0.82  & 39.40 $\pm$ 0.97  & 44.23 $\pm$ 0.43  & 40.53 $\pm$ 1.20  \\
\midrule
\multirow{3}{*}{\C{\hspace{-0.15cm}Test\\ Time\\ (ms)}} &\C{RSPI} & 78.17 $\pm$ 1.43 & 79.83 $\pm$ 3.35 & 85.91 $\pm$ 1.83 & 77.58 $\pm$ 2.29 & 82.89 $\pm$ 1.34 & 83.96 $\pm$ 1.52 & 81.97 $\pm$ 1.71 \\
\cmidrule{2-9}
&\C{NANO} & 39.46 $\pm$ 0.23 & 42.07 $\pm$ 1.89 & 46.81 $\pm$ 3.01 & 44.16 $\pm$ 0.13 & 39.58 $\pm$ 0.88 & 42.80 $\pm$ 1.93 & 43.10 $\pm$ 0.28 \\
\cmidrule{2-9}
&\C{Server} & 11.77 $\pm$ 0.10 & 12.68 $\pm$ 0.37 & 12.57 $\pm$ 0.74 & 12.74 $\pm$ 0.26 & 12.15 $\pm$ 0.31 & 12.27 $\pm$ 0.45 & 13.15 $\pm$ 0.16 \\
\midrule
\multicolumn{2}{|c|}{\C{Space (kB)}} & \C{877.72 $\pm$\\ 3.35} & \C{621.78 $\pm$\\ 1.27} & \C{1601.35 $\pm$\\ 0.26} & \C{640.69 $\pm$\\ 0.13} & \C{621.53 $\pm$\\ 0.19} & \C{902.46 $\pm$\\ 0.83} & \C{600.78 $\pm$\\ 0.10} \\
\bottomrule
\end{tabular}
\end{table*}
\FloatBarrier

\begin{table*}[!hbp]
\VLS
\caption{Retained AUC (method over OCSVM) and server train time speedup
(OCSVM over method) with STATS+HEADER.}
\label{tab:stats_header-best-retained}
\begin{tabular}{|F{2.6cm}|F{1.9cm}|F{1.9cm}|F{1.9cm}|F{1.9cm}|F{1.9cm}|F{1.9cm}|F{1.9cm}|} 
\toprule
\diagbox[width=2.6cm,height=.6cm]{\raisebox{-0.02cm}{\vspace{0.0cm}Method}}{\raisebox{0.5cm}{\vspace{-0.3cm}\hspace{-1.0cm}Dataset}}    & UNB      & CTU  & MAWI         & MACCDC       & SFRIG  & AECHO   &DWSHR  \\
\midrule
\C{\hspace{-1.4cm}OC-KJL:\\\hspace{0.0cm}\uline{AUC Retained}} &1.35 $\pm$ 0.07 & 0.98 $\pm$ 0.10 & 0.99 $\pm$ 0.01 & 0.95 $\pm$ 0.04 & 0.99 $\pm$ 0.01 & 1.01 $\pm$ 0.00 & 0.99 $\pm$ 0.02\\
\cmidrule{2-8}
\C{Train Speedup} &2.45 $\pm$ 0.05  & 1.93 $\pm$ 0.03  & 3.08 $\pm$ 0.07  & 2.34 $\pm$ 0.05  & 2.02 $\pm$ 0.05  & 2.34 $\pm$ 0.02  & 2.06 $\pm$ 0.06 \\
\midrule
\C{\hspace{-0.8cm}OC-KJL-QS:\\\hspace{-0.1cm}\uline{AUC Retained}} &1.30 $\pm$ 0.07 & 1.01 $\pm$ 0.06 & 0.95 $\pm$ 0.04 & 1.10 $\pm$ 0.08 & 0.99 $\pm$ 0.01 & 1.00 $\pm$ 0.02 & 0.99 $\pm$ 0.01\\
\cmidrule{2-8}
\C{Train Speedup} &1.05 $\pm$ 0.02  & 1.07 $\pm$ 0.02  & 1.56 $\pm$ 0.04  & 1.13 $\pm$ 0.02  & 1.07 $\pm$ 0.03  & 1.26 $\pm$ 0.01  & 1.16 $\pm$ 0.03 \\
\midrule
\hline
\midrule
\C{\hspace{-.8cm}OC-Nystr\"om:\\\hspace{0.0cm}\uline{AUC Retained}} &1.44 $\pm$ 0.01 & 1.04 $\pm$ 0.03 & 0.99 $\pm$ 0.00 & 0.94 $\pm$ 0.10 & 0.98 $\pm$ 0.00 & 1.01 $\pm$ 0.00 & 0.98 $\pm$ 0.00\\
\cmidrule{2-8}
\C{Train Speedup} &2.26 $\pm$ 0.05  & 2.13 $\pm$ 0.04  & 3.05 $\pm$ 0.07  & 2.60 $\pm$ 0.05  & 2.09 $\pm$ 0.05  & 2.56 $\pm$ 0.03  & 2.43 $\pm$ 0.07 \\
\midrule
\C{\hspace{-.2cm}OC-Nystr\"om-QS:\\\hspace{0.0cm}\uline{AUC Retained}} &1.42 $\pm$ 0.02 & 0.99 $\pm$ 0.05 & 0.98 $\pm$ 0.01 & 0.85 $\pm$ 0.11 & 0.98 $\pm$ 0.00 & 1.01 $\pm$ 0.00 & 0.97 $\pm$ 0.01\\
\cmidrule{2-8}
\C{Train Speedup} &0.98 $\pm$ 0.02  & 1.07 $\pm$ 0.02  & 1.53 $\pm$ 0.04  & 1.09 $\pm$ 0.02  & 0.97 $\pm$ 0.02  & 1.21 $\pm$ 0.01  & 1.13 $\pm$ 0.03 \\
\bottomrule
\end{tabular}
\end{table*}
\FloatBarrier

\FloatBarrier
\begin{table*}[!htbp]
\VLS
\caption{OC-KJL-QS: Test time speedup
(OCSVM over method) and space reduction (OCSVM over method) with STATS+HEADER.}
\label{tab:stats_header-best-oc_kjl_qs}
\begin{tabular}{|F{1.2cm}|m{1cm}|F{1.9cm}|F{1.9cm}|F{1.9cm}|F{1.9cm}|F{1.9cm}|F{1.9cm}|F{1.9cm}|} 
\toprule
\multicolumn{2}{|c|}{Dataset}    & UNB      & CTU  & MAWI         & MACCDC       & SFRIG  & AECHO   &DWSHR  \\
\midrule
\multirow{3}{*}{\C{\hspace{-0.15cm}Test\\ Time\\Speedup}} &  \C{RSPI} & 18.38 $\pm$ 0.34& 22.62 $\pm$ 0.95& 22.88 $\pm$ 0.49& 21.72 $\pm$ 0.64& 20.03 $\pm$ 0.32& 21.14 $\pm$ 0.38& 23.12 $\pm$ 0.48\\
\cmidrule{2-9}
& \C{NANO} & 9.83 $\pm$ 0.06 & 13.87 $\pm$ 0.62 & 14.57 $\pm$ 0.94 & 14.15 $\pm$ 0.04 & 10.17 $\pm$ 0.23 & 11.77 $\pm$ 0.53 & 13.77 $\pm$ 0.09\\
\cmidrule{2-9}
&\C{Server} & 15.67 $\pm$ 0.14& 20.08 $\pm$ 0.59& 19.46 $\pm$ 1.14& 17.26 $\pm$ 0.36& 15.13 $\pm$ 0.39& 18.63 $\pm$ 0.69& 19.08 $\pm$ 0.23\\
\midrule
\multicolumn{2}{|c|}{\C{Space Reduction}} & 20.29 $\pm$ 0.08 & 20.74 $\pm$ 0.04 & 23.16 $\pm$ 0.00 & 20.64 $\pm$ 0.00 & 19.08 $\pm$ 0.01 & 21.10 $\pm$ 0.02 & 20.38 $\pm$ 0.00\\
\bottomrule
\end{tabular}
\end{table*}
\FloatBarrier

\FloatBarrier
\begin{table*}[!hbp]
\VLS
\caption{OC-Nystr\"om-QS: Test time speedup
(OCSVM over method) and space reduction (OCSVM over method)  with STATS+HEADER.}
\label{tab:stats_header-best-oc_nystrom_qs}
\begin{tabular}{|F{1.2cm}|m{1cm}|F{1.9cm}|F{1.9cm}|F{1.9cm}|F{1.9cm}|F{1.9cm}|F{1.9cm}|F{1.9cm}|} 
\toprule
\multicolumn{2}{|c|}{Dataset}    & UNB      & CTU  & MAWI         & MACCDC       & SFRIG  & AECHO   &DWSHR  \\
\midrule
\multirow{3}{*}{\C{\hspace{-0.15cm}Test\\ Time\\Speedup}} &  \C{RSPI} &18.12 $\pm$ 0.33& 22.03 $\pm$ 0.92& 22.88 $\pm$ 0.49& 20.11 $\pm$ 0.59& 18.97 $\pm$ 0.31& 20.58 $\pm$ 0.37& 22.53 $\pm$ 0.47\\
\cmidrule{2-9}
& \C{NANO} &9.97 $\pm$ 0.06 & 13.51 $\pm$ 0.61 & 14.78 $\pm$ 0.95 & 13.28 $\pm$ 0.04 & 10.25 $\pm$ 0.23 & 12.00 $\pm$ 0.54 & 13.71 $\pm$ 0.09\\
\cmidrule{2-9}
&\C{Server} &13.45 $\pm$ 0.12& 21.62 $\pm$ 0.63& 19.95 $\pm$ 1.17& 17.88 $\pm$ 0.37& 14.84 $\pm$ 0.38& 16.35 $\pm$ 0.60& 19.82 $\pm$ 0.24\\
\midrule
\multicolumn{2}{|c|}{\C{Space Reduction}} &20.28 $\pm$ 0.08 & 20.46 $\pm$ 0.04 & 23.17 $\pm$ 0.00 & 20.15 $\pm$ 0.00 & 18.90 $\pm$ 0.01 & 20.99 $\pm$ 0.02 & 20.33 $\pm$ 0.00\\

\bottomrule
\end{tabular}
\end{table*}
\FloatBarrier

\subsubsection{Results Under No Tuning}
OCSVM results under no tuning, for STATS+HEADER features are presented in Table \ref{tab:stat_header-default-ocsvm_baseline}. As with the case of our preferred features of IAT+SIZE, we observe a significant decrease in AUC w.r.t. the tuned OCSVM case. 

 Table \ref{tab:stats_header-default-retained} shows that OC-Nystr\"om-QS and OC-KJL-QS, retain the AUC and train time of the baseline OCSVM using the STATS+HEADER features. 

We also get similar significant test time speedup and space reduction results for both methods as shown in Tables \ref{tab:stats_header-default-oc_kjl_qs} and  \ref{tab:stats_header-default-oc_nystrom_qs}. This goes to show that the reductions inherent in our approach is likely not tied to feature representations of the networking data.

\FloatBarrier
\begin{table*}[!hbp]
\VLS
\caption{OCSVM performance with STATS+HEADER, \textbf{no tuning}. Time is in ms per 100 datapoints and space is in kB. }
\label{tab:stat_header-default-ocsvm_baseline}
\begin{tabular}{|F{1.2cm}|m{1cm}|F{1.9cm}|F{1.9cm}|F{1.9cm}|F{1.9cm}|F{1.9cm}|F{1.9cm}|F{1.9cm}|} 
\toprule
\multicolumn{2}{|c|}{Dataset}    & UNB      & CTU  & MAWI         & MACCDC       & SFRIG  & AECHO   &DWSHR  \\
\midrule
\multicolumn{2}{|c|}{AUC} &0.49 $\pm$ 0.00 & 0.33 $\pm$ 0.00 & 0.99 $\pm$ 0.00 & 0.44 $\pm$ 0.00 & 0.92 $\pm$ 0.00 & 0.97 $\pm$ 0.00 & 0.63 $\pm$ 0.00 \\
\midrule
\multicolumn{2}{|c|}{\C{Server Train \\Time (ms)}} & 39.13 $\pm$ 0.65  & 40.87 $\pm$ 1.22  & 52.63 $\pm$ 0.91  & 40.86 $\pm$ 0.59  & 38.89 $\pm$ 1.27  & 44.60 $\pm$ 1.42  & 39.73 $\pm$ 1.14  \\
\midrule
\multirow{3}{*}{\C{\hspace{-0.15cm}Test\\ Time\\ (ms)}} &\C{RSPI} & 78.07 $\pm$ 1.16 & 81.06 $\pm$ 1.54 & 83.40 $\pm$ 1.18 & 83.12 $\pm$ 1.38 & 81.83 $\pm$ 1.14 & 82.96 $\pm$ 1.13 & 82.03 $\pm$ 1.12 \\
\cmidrule{2-9}
&\C{NANO}  & 37.97 $\pm$ 0.19 & 44.29 $\pm$ 0.18 & 40.57 $\pm$ 0.10 & 43.08 $\pm$ 0.10 & 40.77 $\pm$ 0.06 & 44.98 $\pm$ 0.18 & 41.60 $\pm$ 0.11 \\
\cmidrule{2-9}
& \C{\hspace{0.cm}Server} & 11.33 $\pm$ 0.07 & 13.26 $\pm$ 0.38 & 11.20 $\pm$ 0.04 & 13.27 $\pm$ 0.13 & 12.63 $\pm$ 0.12 & 12.92 $\pm$ 0.08 & 12.51 $\pm$ 0.27 \\
\midrule
\multicolumn{2}{|c|}{\C{Space (kB)}}& \C{862.03 $\pm$\\ 0.46} & \C{621.43 $\pm$\\ 0.37} & \C{1601.99 $\pm$\\ 0.48} & \C{641.92 $\pm$\\ 0.44} & \C{621.48 $\pm$\\ 0.12} & \C{901.95 $\pm$\\ 0.27} & \C{601.59 $\pm$\\ 0.24} \\
\bottomrule
\end{tabular}
\end{table*}
\FloatBarrier

\FloatBarrier
\begin{table*}[!htbp]
\VLS
\caption{\textbf{no tuning}. Retained AUC (method over OCSVM) and train time speedup
(OCSVM over method) with STATS+HEADER.}
\label{tab:stats_header-default-retained}
\begin{tabular}{|F{2.5cm}|F{1.9cm}|F{1.9cm}|F{1.9cm}|F{1.9cm}|F{1.9cm}|F{1.9cm}|F{1.9cm}|} 
\toprule
\diagbox[width=2.6cm,height=.6cm]{\raisebox{-0.02cm}{\vspace{0.0cm}Method}}{\raisebox{0.5cm}{\vspace{-0.3cm}\hspace{-1.0cm}Dataset}}    & UNB      & CTU  & MAWI         & MACCDC       & SFRIG  & AECHO   &DWSHR  \\
\midrule
\C{\hspace{-.7cm}OC-KJL-QS:\\\hspace{-0.cm}\uline{AUC Retained}} &1.66 $\pm$ 0.05 & 1.51 $\pm$ 0.07 & 0.09 $\pm$ 0.01 & 1.16 $\pm$ 0.33 & 0.89 $\pm$ 0.05 & 0.82 $\pm$ 0.08 & 0.88 $\pm$ 0.04\\
\cmidrule{2-8}
\C{Train Speedup}  &0.93 $\pm$ 0.02  & 1.13 $\pm$ 0.03  & 1.29 $\pm$ 0.02  & 1.15 $\pm$ 0.02  & 1.06 $\pm$ 0.03  & 1.25 $\pm$ 0.04  & 1.09 $\pm$ 0.03 \\
\midrule
\C{\hspace{-0.1cm}OC-Nystr\"om-QS:\\\hspace{-0.cm}\uline{AUC Retained}} &1.69 $\pm$ 0.06 & 1.77 $\pm$ 0.08 & 0.11 $\pm$ 0.00 & 1.40 $\pm$ 0.15 & 0.35 $\pm$ 0.27 & 0.17 $\pm$ 0.04 & 0.67 $\pm$ 0.02\\
\cmidrule{2-8}
\C{Train Speedup} &0.94 $\pm$ 0.02  & 1.12 $\pm$ 0.03  & 1.11 $\pm$ 0.02  & 1.13 $\pm$ 0.02  & 1.04 $\pm$ 0.03  & 1.19 $\pm$ 0.04  & 1.09 $\pm$ 0.03 \\
\bottomrule
\end{tabular}
\end{table*}
\FloatBarrier

\FloatBarrier
\begin{table*}[!htbp]
\VLS
\caption{OC-KJL-QS with STATS+HEADER, \textbf{no tuning}: Test time speedup
(OCSVM over method) and space reduction (OCSVM over method).}
\label{tab:stats_header-default-oc_kjl_qs}
\begin{tabular}{|F{1.2cm}|m{1.cm}|F{1.9cm}|F{1.9cm}|F{1.9cm}|F{1.9cm}|F{1.9cm}|F{1.9cm}|F{1.9cm}|} 
\toprule
\multicolumn{2}{|c|}{Dataset}    & UNB      & CTU  & MAWI         & MACCDC       & SFRIG  & AECHO   &DWSHR  \\
\midrule
\multirow{3}{*}{\C{\hspace{-0.1cm}Test\\ Time\\Speedup}} &  RSPI & 17.30 $\pm$ 0.26& 22.71 $\pm$ 0.43& 17.27 $\pm$ 0.24& 22.88 $\pm$ 0.38& 19.22 $\pm$ 0.27& 20.18 $\pm$ 0.27& 23.41 $\pm$ 0.32\\
\cmidrule{2-9}
& \C{NANO} & 9.57 $\pm$ 0.05 & 13.99 $\pm$ 0.06 & 10.16 $\pm$ 0.03 & 13.75 $\pm$ 0.03 & 11.45 $\pm$ 0.02 & 12.59 $\pm$ 0.05 & 14.19 $\pm$ 0.04\\
\cmidrule{2-9}
&\C{Server} & 14.71 $\pm$ 0.09& 20.75 $\pm$ 0.60& 13.74 $\pm$ 0.05& 23.26 $\pm$ 0.23& 18.40 $\pm$ 0.18& 17.36 $\pm$ 0.11& 20.46 $\pm$ 0.44\\
\midrule
\multicolumn{2}{|c|}{\C{Space Reduction}} & 19.93 $\pm$ 0.01 & 20.66 $\pm$ 0.01 & 21.99 $\pm$ 0.01 & 20.68 $\pm$ 0.01 & 19.29 $\pm$ 0.00 & 21.11 $\pm$ 0.01 & 20.61 $\pm$ 0.01\\

\bottomrule
\end{tabular}
\end{table*}
\FloatBarrier

\FloatBarrier
\begin{table*}[!htbp]
\VLS
\caption{OC-Nystr\"om-QS with STATS+HEADER, \textbf{no tuning}: Test time speedup
(OCSVM over method) and space reduction (OCSVM over method).}
\label{tab:stats_header-default-oc_nystrom_qs}
\begin{tabular}{|F{1.2cm}|m{1cm}|F{1.9cm}|F{1.9cm}|F{1.9cm}|F{1.9cm}|F{1.9cm}|F{1.9cm}|F{1.9cm}|} 
\toprule
\multicolumn{2}{|c|}{Dataset}    & UNB      & CTU  & MAWI         & MACCDC       & SFRIG  & AECHO   &DWSHR  \\
\midrule
\multirow{3}{*}{\C{\hspace{-0.15cm}Test\\ Time\\Speedup}} &  \C{RSPI} &17.26 $\pm$ 0.26& 22.70 $\pm$ 0.43& 17.16 $\pm$ 0.24& 21.44 $\pm$ 0.36& 18.60 $\pm$ 0.26& 19.08 $\pm$ 0.26& 22.06 $\pm$ 0.30\\
\cmidrule{2-9}
& \C{NANO} &9.66 $\pm$ 0.05 & 14.19 $\pm$ 0.06 & 9.88 $\pm$ 0.03 & 12.88 $\pm$ 0.03 & 11.13 $\pm$ 0.02 & 12.29 $\pm$ 0.05 & 13.58 $\pm$ 0.04\\
\cmidrule{2-9}
&\C{Server} &13.76 $\pm$ 0.09& 24.66 $\pm$ 0.71& 14.86 $\pm$ 0.06& 20.94 $\pm$ 0.20& 16.42 $\pm$ 0.16& 16.31 $\pm$ 0.10& 18.85 $\pm$ 0.40\\
\midrule
\multicolumn{2}{|c|}{\C{Space Reduction}} &19.92 $\pm$ 0.01 & 20.58 $\pm$ 0.01 & 21.98 $\pm$ 0.01 & 20.13 $\pm$ 0.01 & 19.02 $\pm$ 0.00 & 20.84 $\pm$ 0.01 & 20.29 $\pm$ 0.01\\
\bottomrule
\end{tabular}
\end{table*}
\FloatBarrier

\end{document}